\documentclass[fleqn,usenatbib]{mnras}

\usepackage[T1]{fontenc}
\usepackage{times}

\DeclareRobustCommand{\VAN}[3]{#2}
\let\VANthebibliography\thebibliography
\def\thebibliography{\DeclareRobustCommand{\VAN}[3]{##3}\VANthebibliography}

\usepackage{graphicx}	
\usepackage{amsmath}	
\usepackage{amssymb}	

\usepackage[dvipsnames]{xcolor}    

\newcommand{\gramses}{\textsc{gramses}}
\newcommand{\dtfe}{\textsc{dtfe}}
\newcommand{\camb}{\textsc{camb}}

\def\Mpch{~h^{-1} {\rm Mpc}}
\def\invMpch{~h{\rm Mpc}^{-1}}
\def\kpch{~h^{-1} {\rm kpc}}

\def\Msolar{~h^{-1}{M}_{\odot}}

\title[Nonlinear vector modes from \gramses{} simulations]{Vector modes in $\Lambda$CDM: the gravitomagnetic potential in dark matter haloes from relativistic $N$-body simulations}

\author[C. Barrera-Hinojosa {\it et al.}]{
Cristian Barrera-Hinojosa,$^{1}$\thanks{E-mail: cristian.g.barrera@durham.ac.uk}
Baojiu Li$^{1}$
, Marco Bruni$^{2,3}$ and Jian-hua He$^{4,5}$
\\
$^{1}$Institute for Computational Cosmology, Department of Physics, Durham University,
Durham DH1 3LE, UK\\
$^{2}$Institute of Cosmology and Gravitation, University of Portsmouth, Dennis Sciama Building, Burnaby Road, Portsmouth PO1 3FX, UK\\
$^{3}$INFN Sezione di Trieste, Via Valerio 2, 34127 Trieste, Italy\\
$^{4}$School of Astronomy and Space Science,  Nanjing University, Nanjing 210023, China \\
$^{5}$Key Laboratory of Modern Astronomy and Astrophysics (Nanjing University), Ministry of
Education, Nanjing 210023, China
}

\date{Accepted XXX. Received YYY; in original form ZZZ}

\pubyear{2020}

\begin{document}
\label{firstpage}
\pagerange{\pageref{firstpage}--\pageref{lastpage}}
\maketitle

\begin{abstract}
We investigate the transverse modes of the gravitational and velocity fields in $\Lambda$CDM, based on a high-resolution simulation performed using the adaptive-mesh refinement general-relativistic $N$-body code \gramses. We study the generation of vorticity in the dark matter velocity field at low redshift, providing fits to the shape and evolution of its power spectrum over a range of scales. By analysing the gravitomagnetic vector potential, which is absent in Newtonian simulations, in dark matter haloes with masses ranging from $\sim10^{12.5}\Msolar$ to $\sim10^{15}\Msolar$, we find that its magnitude correlates with the halo mass, peaking in the inner regions. Nevertheless, on average, its ratio against the scalar gravitational potential remains fairly constant, below percent level, decreasing roughly linearly with redshift and showing a weak dependence on halo mass. Furthermore, we show that the gravitomagnetic acceleration in haloes peaks towards the core and reaches almost $10^{-10}$ $h$ cm/s$^2$ in the most massive halo of the simulation. 
However, regardless of the halo mass, the ratio between the gravitomagnetic force and the standard gravitational force is typically at around the 
$10^{-5}$ level inside the haloes, again without significant radius dependence.
This result confirms that the gravitomagnetic effects have negligible impact on structure formation, even for the most massive structures, although its behaviour in low density regions remains to be explored. Likewise, the impact on observations remains to be understood in the future.

\end{abstract}




\section{Introduction}

While the dynamics of the large-scale structure (LSS) of the universe is mainly governed by scalar perturbations, vector and tensor degrees of freedom are promising alternatives for exploring the nature of dark matter and gravity. 
The effects of the vector modes of the spacetime metric on matter such as frame dragging and geodetic precession have been measured in the Solar system during the last decade~\citep{GravityProbeB:2011}, but there is still no cosmological signal detected. The recent observation of radio galaxies showing coherent angular velocities on scales of $\sim20$ Mpc at $z=1$ reported by~\citet{Taylor_2016} has motivated to seek a physical interpretation in terms of vector modes, but it has not been possible to establish a clear connection so far~\citep{cusin_vorticity_2017,bonvin_redshift-space_2018}. More recently, and motivated by the accurate data provided by {\it Gaia} DR2, a simple model to explain the flat rotation curve of the Milky Way in terms of frame dragging has been proposed in~\citet{Crosta:2020}. 

In $\Lambda$CDM cosmology, vector modes are typically neglected. 
In a perfect fluid,  { vorticity -- the covariant curl of the 4-velocity field -- satisfies a homogenous nonlinear equation, hence it vanishes exactly, i.e.\ at all orders in perturbation theory \citep{Lu:2008ju}, unless
 it is either introduced by initial conditions\footnote{Even if non zero initially, during expansion a first-order vorticity in a standard perfect fluid is red-shifted away. } or generated by physics beyond such fluid model}. Moreover, vorticity is not generated by standard inflationary scenarios, and even if it was, this type of perturbation quickly decays during the matter-dominated era. Nonetheless, vorticity is found to be generated dynamically via shell (orbit) crossing of 
matter, 
a phenomenon extremely common at late times whose modelling is beyond the grasp of the single-streaming fluid regime. Therefore, $N$-body simulations represent a valuable tool for the study of vorticity generation \citep{pueblas_generation_2009,hahn_properties_2015,jelic-cizmek_generation_2018}.

In the  Poisson gauge, generalising the longitudinal gauge to include tensor and vector perturbations \citep{Bertschinger:1993xt},  the latter  are encoded by the non-diagonal spacetime metric components, the shift vector $B_i\equiv g_{0i}$, and represent in this gauge the gauge-invariant gravitomagnetic vector potential \citep{Bardeen:1980kt}. In $\Lambda$CDM, {safely assuming purely scalar perturbations at first-order,  the shift vector} vanishes at the linear level, while at second order it satisfies a constraint equation sourced by the product of first-order density and velocity perturbations. However, it is expected that, just like vorticity, the {gravito-magnetic field} also receives corrections from phenomena beyond the perfect fluid description. 

The impact of vector modes on LSS observables is expected to be small relative to the scalar perturbations, {both from  perturbative \citep{Lu:2008ju} and non-perturbative analyses \citep{bruni_computing_2014,gevolution-main},} although it can represent a new systematic which needs to be taken into account \citep{bonvin_redshift-space_2018}.
For instance, their effect on gravitational lensing seems to be not strong enough to be detectable by current observations \citep{Thomas:2014aga,Saga:2015apa,Gressel:2019jxw}, and the imprints of the vector potential in the angular power spectrum and bispectrum of galaxies are also weak \citep{durrer_vector_2016,jolicoeur_imprints_2019}, although a vector perturbation can be isolated from the full signal if it violates statistical isotropy and defines a preferred frame \citep[see, e.g.,][]{tansella_redshift-space_2018}. 
On the other hand, the vector potential power spectrum is known to peak around the equality scale \citep{Lu:2008ju}, and its behaviour as well as impact on observables at {highly} nonlinear scales remains largely unexplored, although deviations from perturbation theory can be significant \citep{bruni_computing_2014}. Furthermore, in popular $f(R)$ gravity models, vector modes can have considerable deviations from GR on small scales \citep{thomas_fr_2015}, so these could also play a role in discriminating cosmological models. 

The work of \citet{pueblas_generation_2009} provided the first insights into the generation of vorticity via shell crossing using $N$-body simulations, which allowed to quantify its impact on the density and velocity power spectra estimates from linear perturbation theory. In particular, vorticity was found to peak in the outskirts of virialised structures as particle velocities in inner regions are strongly aligned with density gradients, as also found later in \citet{hahn_properties_2015} from a different set of simulations. Although -- contrary to vorticity -- the investigation of the {gravitomagnetic vector field} in principle requires a completely general-relativistic numerical framework as Newtonian simulations only model a single scalar gravitational potential, $\Phi$, in \citet{bruni_computing_2014} and \citet{thomas_fully_2015} a novel method to extract its power spectrum by post-processing the momentum density field from a Newtonian simulation was introduced. This is motivated by the fact that the leading contribution to the shift vector in post-Friedmann expansion \citep{Milillo:2015cva} is sourced by the transverse part of the momentum density field. Although this method neglects the feedback of the shift vector into the simulation dynamics, this approximation is well justified as perturbation theory estimates that the magnitude of the vector potential is at most one percent of the scalar gravitational potential \citep{Lu:2008ju}.

Cosmological codes which are capable of simulating vector modes of the metric have been only recently developed \citep[e.g.,][]{Adamek:2016zes,gevolution-main,Mertens:2015ttp,Giblin:2017juu,Macpherson:2016ict,gramses-1}, and have proven robust enough to study different relativistic distortions in the large-scale structures (LSS); \citep[see][for an actual comparison of frame-dragging observables in a toy universe simulated using these codes]{Adamek:2020jmr}. In particular, the cross correlation between the shift vector and vorticity has been studied in \citet{jelic-cizmek_generation_2018} using the relativistic $N$-body code {\it gevolution} \citep{Adamek:2016zes,gevolution-main}, showing that the vector potential is only weakly sourced by vorticity alone, which is subdominant 
compared with the density-dependent terms coming from the transverse projection of the {\it full} momentum field, in qualitative agreement with post-Friedmann expansion results from \citet{bruni_computing_2014,thomas_fully_2015}. 

The objective of this paper is to study the vector modes of both the gravitational and matter velocity fields from large sub-horizon scales down to deeply nonlinear scales using \gramses\ \citep{gramses-1,gramses-2}, a recently-introduced {general-}relativistic $N$-body code based on {\sc ramses} \citep{Teyssier:ramses}. We expand on previous studies in the following ways: (i) similarly to~\citet{jelic-cizmek_generation_2018}, we provide a direct calculation of the {gravitomagnetic field, represented by the  shift vector,} from the simulation, also relaxing the weak-field approximation in our approach; (ii) we present results for scales in the deeply nonlinear regime which have not been previously explored in this context, and which are accessible thanks to the adaptive-mesh refinement (AMR) capabilities of \gramses. For the first time, we explore the gravitomagnetic vector potential in dark matter haloes in a broad range of halo masses; (iii) furthermore, we quantify the gravitomagnetic acceleration inside the dark matter haloes and compare this against the standard gravitational one.

We note that, with the exception of \citet{jelic-cizmek_generation_2018}, previous studies of vorticity use simulations that incorporate a softening length scale, a numerical parameter used to prevent divergences in the calculation of inter-particle forces which also determines the spatial resolution. In \gramses\ -- similarly to {\it gevolution} -- the metric components and their spatial derivatives are calculated on a Cartesian mesh. AMR codes, such as \gramses, are generally slower than fixed-mesh-resolution codes such as {\it gevolution} which can benefit from efficient standard libraries such as {\sc fftw}, but their adaptively-produced mesh structure in high-density regions allows them to be more focused on the fine details in such regions, without increasing the overall cost of the simulation substantially. Therefore, they provide complementary ways to study the vector modes from cosmological simulations.

The rest of this paper is organised as follows. In Section~\ref{sec:method} we fix our notations and briefly describe the general-relativistic formalism and methods implemented in the \gramses~code that are relevant for the vector modes. In Section~\ref{sec:spectra} we show the results for the different power spectra of the velocity field components as well as of the gravitomagnetic potential. Then, in Section~\ref{sec:haloes}, we focus on dark matter haloes, providing comparisons of the gravitomagnetic potential and corresponding acceleration with the scalar counterparts.

Throughout this paper, Greek indices are used to label spacetime vectors and run over $(0,1,2,3)$, while Latin indices run over $(1,2,3)$. Unless otherwise stated, we follow the unit convention that the speed of light $c=1$.

\section{Method and definitions}
\label{sec:method}

For the sake of clarity and completeness, let us briefly summarise the terminology and conventions adopted in this paper, which in some part stem from \gramses\,' implementation itself. More details can be found in the main code paper \citep{gramses-1} and the references therein. 

In order to solve the gravitational sector equations and geodesic equations, \gramses\ adopts the $3+1$ formalism in which the spacetime metric takes the form
\begin{equation}\label{eq:adm-metric}
{\rm d}s^2 = g_{\mu\nu}{\rm d}x^{\mu}{\rm d}x^{\nu} = -\alpha^2 {\rm d}t^2+\gamma_{ij}\left(\beta^i{\rm d}t+{\rm d}x^i\right)\left(\beta^j{\rm d}t+{\rm d}x^j\right),
\end{equation}
where $\alpha$ is the lapse function, $\beta^i$ the shift vector
and $\gamma_{ij}$ the induced metric on the spatial hypersurfaces, which in the constrained formulation adopted by \gramses\ is approximated by a conformally-flat metric, $\gamma_{ij}=\psi^4\delta_{ij}$, with $\psi$ being the conformal factor and $\delta_{ij}$ the Kronecker delta. 

In the $3+1$ formalism $n_\mu=(-\alpha,0)$ is the unit timelike vector normal to the time slices, the 3-dimensional spatial hypersurfaces with metric $\gamma_{ij}$, and   Eulerian observers are those with 4-velocity $n^\mu$. 
The energy density $\rho$ and momentum density $S_i$ measured by these  normal observers are given by the following projections of the energy-momentum tensor $T^{\mu\nu}$,
\begin{align}
\rho &\equiv n_\mu n_\nu T^{\mu\nu}\,,\label{rho-def}\\
S_i&\equiv-\gamma_{i\mu}n_\nu T^{\mu\nu}\,,\label{S_i-def}
\end{align}
where the action of $n_\mu$ projects onto the timelike direction, while $\gamma_{\mu\nu}=g_{\mu\nu}+n_\mu n_\nu$ projects onto the spatial hypersurface.
Eq.~\eqref{rho-def} and \eqref{S_i-def} are the source terms for the Hamiltonian constraint and momentum constraint, respectively. Additionally, the spatial stress and its trace are defined as
\begin{align}
    S_{ij}&\equiv\gamma_{i\mu}\gamma_{j\nu}T^{\mu\nu}\,, \qquad\qquad S=\gamma^{ij}S_{ij}\,,
\end{align}
which, in addition to $\rho$ and $S_i$, appear in the evolution equations for the extrinsic curvature tensor. In \gramses, the (dark) matter sector is represented by an ensemble of non-interacting simulation particles of rest mass $m$ and four-velocity $u^\mu=dx^\mu/d\tau$, where $\tau$ is an affine parameter. 
The equations for the gravitational sector are numerically solved based on conformal matter sources, which are scaled using $\gamma=\det(\gamma_{ij})$ as
\begin{alignat}{2}
s_0({\bf x})&\equiv\sqrt{\gamma}\rho&&\propto{m\alpha{u}^0}\,,\label{eq:s0-cic}\\
s_i({\bf x})&\equiv\sqrt{\gamma}S_i&&\propto m{u}_i\,,\\
s_{ij}({\bf x})&\equiv\sqrt{\gamma}S_{ij}&&\propto m\frac{{u}_i{u}_j}{\alpha{u}^0}\,.\label{eq:s-cic}
\end{alignat}
In these, ${\bf x}$ is a (discrete) position vector on the cartesian simulation grid and the proportionality symbol in each equation stands for the standard cloud-in-cell (CIC) weights used for the particle-mesh projection \citep{CIC-book}. From Eqs.~\eqref{eq:s0-cic}-\eqref{eq:s-cic} we have the following useful relations:
\begin{align}
    s_0&=\rho\Gamma\,,\\
    s_i&=\frac{\rho}{\Gamma}u_i\,,\label{eq:s_i-rel}\\
    s_{ij}&=\frac{\rho}{\Gamma^2}u_iu_j\,, \qquad\implies s=\rho(1-\Gamma^{-2})\,,\\
    u_i&=\Gamma^2\frac{s_i}{s_0}\,,
\end{align}
where $\Gamma\equiv\alpha{u}^0=\sqrt{1+\gamma^{ij}u_iu_j}$ is the Lorentz factor. For a perfect fluid, $s\equiv\sqrt{\gamma}S$ is proportional to pressure in linear theory, and then it vanishes for CDM (dust) in such regime. Naturally, $s$ also vanishes in the non-relativistic limit.

The equations of motion for collisionless particles correspond to the geodesic equation $u^\mu\nabla_\mu u_\nu=0$, which in the $3+1$ form reads
\begin{align}
    \frac{du_i}{dt}&=-\Gamma\partial_i\alpha+u_j\partial_i\beta^j-\alpha\frac{u_ju_k}{2\Gamma}\partial_i\gamma^{jk}\,,\label{eq:geodesic-eq-1}\\
    \frac{dx^i}{dt}&=\alpha\frac{\gamma^{ij}u_j}{\Gamma}-\beta^i.\label{eq:geodesic-eq-2}
\end{align}
In Eq.~\eqref{eq:geodesic-eq-1}, the term $u_j\partial_i\beta^j$ corresponds to a force that is absent in both the Newtonian limit and the linear perturbation regime. {In the case where $\beta^j$ is purely a vector-type perturbation (e.g., the Poisson gauge), this force term is known as {\it gravitomagnetic force}, in formal analogy with the magnetic Lorentz force.}

\subsection{Vector decomposition}

Given that in this paper we are particularly interested in vector modes (transverse modes), we start by splitting a vector field $V^i$ (${\bf V}$) as 
\begin{align}
    {\bf V} &= {\bf V}_{\parallel} + {\bf V}_\perp\,, \label{eq:S-V-decomp}
\end{align}
where ${\bf V}_{\parallel}$ and ${\bf V}_\perp$ are respectively the scalar (irrotational) and vector (rotational) components, i.e., these satisfy
\begin{align}
     \boldsymbol{\nabla}\times{\bf V}_{\parallel}=\boldsymbol{0},\quad \boldsymbol{\nabla}\cdot {\bf V}_\perp=0\,.\label{eq:curl_div_V}
\end{align}
In the case of the velocity field\footnote{We use ${\bf u}$ to represent the velocity $u_i$ rather than $u^i$, as it is the former that is used in the $3+1$ form of the geodesic equations \eqref{eq:geodesic-eq-1} and \eqref{eq:geodesic-eq-2} which are implemented in \gramses. 
{$u_i$ is what we call `CMC-MD-gauge velocity', and is different from $u^i$.} See \citet{gramses-2} for more details.} $u_i$ (${\bf u}$), we define the velocity divergence and vorticity as 
\begin{align}
    \theta &\equiv \boldsymbol{\nabla}\cdot{\bf u}\,,\\
    \boldsymbol{\omega} &\equiv \boldsymbol{\nabla}\times{\bf u}\,.
\end{align}
As usual, the power spectra of these quantities are respectively defined as
\begin{align}
   \left\langle{\theta}({\bf k)}{\theta}^{*}({\bf k}')\right\rangle&=\delta({\bf k}-{\bf k}')(2\pi)^3P_{\theta}(k)\,,\label{eq:Pk-sca-def}\\
   \left\langle{\boldsymbol{\omega}}^i({\bf k)}{\boldsymbol{\omega}}^{*j}({\bf k}')\right\rangle&=\delta({\bf k}-{\bf k}')(2\pi)^3\frac{1}{2}\left(\delta^{ij}-\frac{k^ik^j}{k^2}\right)P_{\boldsymbol{\omega}}(k)\,,\label{eq:Pk-vec-def}
\end{align}
and the velocity power spectrum satisfies the relation
\begin{equation}
    P_{\bf |u|}=k^2(P_{\theta}+P_{\boldsymbol{\omega}})\,.
\end{equation}
The power spectrum of the vector modes of the shift vector is defined in analogous way to Eq.~\eqref{eq:Pk-vec-def}.

\subsection{Gauge choice and the constraint for the vector potential}

For solving the gravitational and geodesic equations, \gramses{} implements a constrained formulation of GR~\citep{Bonazzola-FCF:2004,CorderoCarrion:2008nf}, in which both the tensor modes of the spatial metric and the transverse-traceless (TT) part of the extrinsic curvature are neglected during the evolution. In contrast, the scalar and vector modes of the gravitational field are treated fully nonlinearly.
In order to do this in a robust way, the formalism adopts the constant-mean-curvature slicing~\citep{Smarr-York:MEC-1978,Shibata:1999va,Shibata:1999zs} and a minimal-distortion gauge condition under the conformal flatness approximation~\citep{Smarr1978:MDC}. Contrary to the Poisson gauge, in this gauge the shift vector contains both scalar and vector ($1+2$) degrees of freedom. At linear order, the latter modes match the gauge-invariant shift vector from the Poisson gauge \citep{matarrese_relativistic_1998,Lu:2008ju}, while the mismatch in the scalar piece reflects the fact that the time foliations are different in these two gauges. 
Then, in this formalism the components of the shift vector are solved from a combination of the $3+1$ evolution equation for the extrinsic curvature, the momentum constraint and the gauge conditions~\citep{gramses-1} 
\begin{align}
(\Delta_L\beta)^i
&=16\pi\alpha\psi^{-6}s^i+\partial_j(\alpha\psi^{-6})\bar{A}^{ij}_L\,,\label{eq:MDC-CF}
\end{align}
where $s^i=\delta^{ij}s_i$, $(\Delta_L \beta)^i:=\partial^2\beta^i+\partial^i(\partial_j\beta^j)/3$ denotes the flat-space vector Laplacian operator, and
\begin{equation}
\bar{A}^{ij}_L=\partial^{i}W^{j}+\partial^{j}W^{i}-\frac{2}{3}\delta^{ij}\partial_kW^k\,,\label{eq:Aij-L-def}
\end{equation}
is the longitudinal part of the traceless extrinsic curvature tensor. The auxiliary potential $W_i$ introduced in Eq.~\eqref{eq:Aij-L-def} is directly solved from the momentum constraint equation,
\begin{equation}\label{eq:W_i-constraint}
(\Delta_L W)_i=16\pi s_i\,.
\end{equation}
Then, from Eq.~\eqref{eq:MDC-CF} we note that, at leading order, the shift vector is sourced by the momentum field and thus $\beta^i\propto W^i$ by Eq.~\eqref{eq:W_i-constraint}, while differences appear at higher order due to the extrinsic curvature tensor sourcing $\beta^i$. 
Given that throughout this paper we will be interested in the vector modes of the shift vector, this is decomposed in the same fashion of Eq.~\eqref{eq:S-V-decomp}, i.e.
\begin{equation}
    \beta^i = {B}^i+\beta^i_\parallel\,,
\end{equation}
where $B^i\equiv\beta^i_\perp$ (${\bf B}$) is referred to as the vector potential or gravitomagnetic potential, and $\beta^i_\parallel$ is the scalar mode of the shift. 
Let us note that, using Eq.~\eqref{eq:s_i-rel}, the curl of the conformal momentum density field $s_i$ (${\bf s}$) can be written non-perturbatively as 
\begin{align}
    \boldsymbol{\nabla}\times{\bf s}=\Gamma^{-1}[(1+\delta){\boldsymbol{\omega}}+\boldsymbol{\nabla}\delta\times{\bf u}-\boldsymbol{\nabla}\Gamma\times{\bf s}]\,,\label{eq:curl-s_i}
\end{align}
where $\delta=\rho/\bar{\rho}-1$ is the density contrast and $\bar{\rho}$ is the mean density. Previous studies have shown that the terms $\delta\boldsymbol{\omega}$ and $\boldsymbol{\nabla}\delta\times{\bf u}$ in the r.h.s. of Eq.~\eqref{eq:curl-s_i} are the main sources for the vector potential \citep{bruni_computing_2014,thomas_fully_2015,jelic-cizmek_generation_2018}, while the contribution from vorticity itself is subdominant at all scales. 
{In the r.h.s. of Eq.~\eqref{eq:curl-s_i}, the last term and the overall modulation by the Lorentz Factor $\Gamma$ arise due to the definition of {\bf s} in Eq.~\eqref{eq:s_i-rel}, 
and both contributions vanish in the linear regime and the non-relativistic limit.}

\section{Results}\label{sec:results}
For the investigation in this paper, we have run a high-resolution simulation using 
\gramses, with a comoving box size $L_{\rm box}=256\Mpch$ and $N_{\rm part}=1024^3$ dark-matter particles, corresponding to a particle mass resolution of $1.33\times10^{9}\Msolar$. Because \gramses~makes use of AMR in high-density regions, the spatial resolution is not uniform throughout the simulation volume: while the coarsest (domain) grid has 
$N_{\rm part}$ cells, corresponding to a comoving spatial resolution of $0.25\Mpch$, the most refined (high density) regions reach a resolution of 
$128^3\times N_{\rm part}$ grid elements, with corresponding spatial resolution of $2\kpch$. 

Initial conditions suitable for the relativistic simulation were generated at $z=49$ with a modified version of {\sc 2lpt}ic code~\citep{Crocce2006:2LPT} fed with the matter power spectrum obtained from a modified version of \camb~\citep{CAMB} that works for the particular gauge needed for \gramses. More details on this can be found in~\citet{gramses-2}. The cosmological parameters adopted for the simulation are $\{\Omega_\Lambda$, $\Omega_m$, $\Omega_K$, $h\}=\{0.693,0.307,0,0.68\}$ and a primordial spectrum with amplitude $A_s=2.1\times10^{-9}$, spectral index $n_s=0.96$ and a pivot scale $k_{\rm pivot}=0.05~{\rm Mpc}^{-1}$. 

In order to measure the velocity fields from simulation snapshots, we use the publicly-available \dtfe~code~\citep{Cautun:2011-DTFE} which is based on the Delaunay tessellation method, although other methods have been explored in the literature during the last few years. Notably, the phase-interpolation method introduced in~\citet{Abel_2012} shows better performance than \dtfe~in shell-crossing regions~\citep{hahn_properties_2015}, where the finite-difference estimation of velocity divergence and vorticity across caustics can be problematic due to the multiply-valued nature of the velocity field. Nonetheless, the 
{power spectra} of these two fields are not {strongly} affected by this since the volume-weighted contribution from caustics is negligible, and both methods converge when nonlinear scales are well resolved. 
In addition, while the vorticity power spectrum is affected by resolution effects, this is weakly affected by finite-volume effects~\citep{pueblas_generation_2009,jelic-cizmek_generation_2018}.
We note that, while the initial velocity field is vorticity-free by construction, spurious vorticity will be present at some degree due to the numerical errors introduced by particle-mesh projections. In addition, shell-crossing events -- which source vorticity -- are rare at high redshift, and its insufficient sampling restricts the possibility of estimating the velocity field robustly. Therefore, in this paper we shall focus mainly on low redshifts, $z<1.5$, at which vorticity results are expected to be robust. 
{Contrary to the velocity field, the gravitomagnetic potential is already solved by the code on a Cartesian mesh so there is no need for post-processing particle-mesh projections.}

It is worthwhile to mention that, although the GR
simulations 
do not necessitate the specification of a cosmological background \citep{gramses-1}, throughout this paper the notion of redshift is still used and should be understood as the standard, background one. This is achieved through the constant-mean-curvature slicing condition, which allows us to fix the trace of the extrinsic curvature of the spatial hypersurfaces as $K=-3H(t)$, where {the Hubble parameter} $H$ can be conveniently fixed via `fiducial' Friedmann equations~\citep{Giblin:2018ndw,gramses-1}.
{In addition, even though in the gauge adopted by \gramses{} the scalar gravitational potentials as well as the matter fields are not gauge-invariant quantities}, gauge effects are only prominent on large scales and become strongly suppressed for modes inside the horizon. Since in this work we are mainly interested in the latter, as well as in redshifts below $z=1.5$ {(in which the horizon is already larger than the box size),}
we do not explore potential gauge issues further. 

\begin{figure*}
    \centering
    \includegraphics[width=1\linewidth]{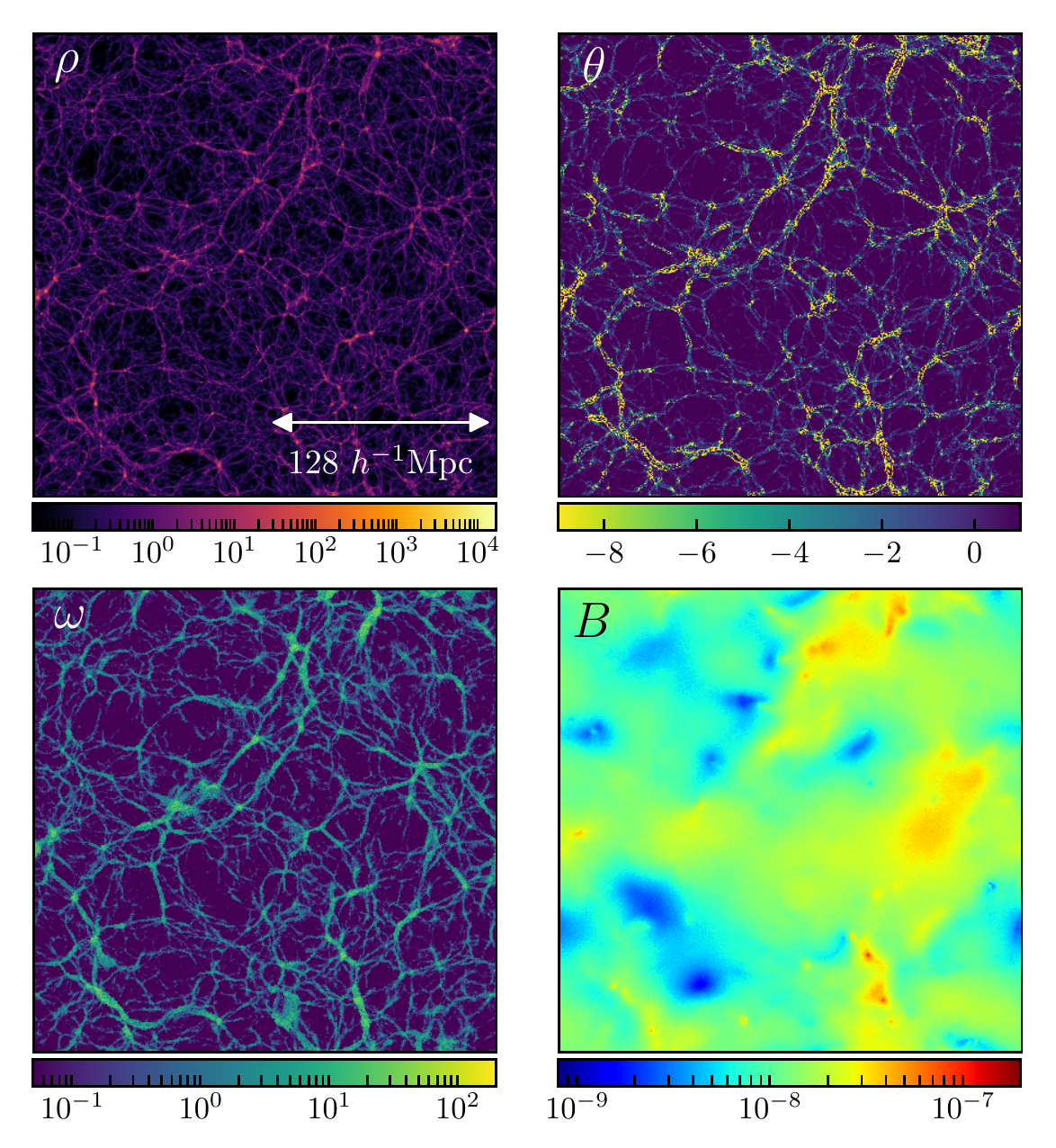}
    \caption{(Colour Online) A slice of the simulation box at $z=0$ showing the density (top left), velocity divergence (top right), vorticity (bottom left) and vector potential magnitude (bottom right) fields. The velocity values shown are normalised by $\mathcal{H}f$, where $\mathcal{H}\equiv aH$ is the conformal Hubble parameter and $f$ is the linear growth rate in $\Lambda$CDM. The density field is normalised by its mean value in the simulation box. }
    \label{fig:maps}
\end{figure*}

Figure~\ref{fig:maps} provides a visual representation of the density field (top left), velocity divergence (top right), the magnitude\footnote{For this we use the flat metric $\delta_{ij}$.} of the vorticity vector field, $\omega\equiv|\boldsymbol{\omega}|=({\omega^2_x+\omega^2_y+\omega^2_z})^{1/2}$ (bottom left), and the vector potential magnitude, $B\equiv|{\bf B}|=({B^2_x+B^2_y+B^2_z})^{1/2}$ (bottom right), across a slice of the simulation box at $z=0$.
From this figure, it is clear that the density field has a similar large-scale distribution to the velocity divergence, consistently with linear perturbation theory. Since velocity divergence can take negative values, we use a linear scale on its map, with a cutoff of extreme values to help visualisation. As expected, the velocity divergence is negative in collapsing regions due to matter in-fall, and positive in voids and low-density regions. The vorticity field also shows a clear correlation with both density and velocity divergence. 
However, we should bear in mind that, as we have discussed before, the velocity divergence and vorticity estimated by \dtfe\ are not completely reliable near caustics \citep{hahn_properties_2015}, and therefore such maps only provide qualitative information and an accurate picture on large scales.

From the bottom right panel in Fig.~\ref{fig:maps}, we observe that the magnitude of the vector potential has some degree of correlation with the structures observed in density, velocity divergence and vorticity, particularly in very high-density and low-density regions. As shown by Eqs.~\eqref{eq:MDC-CF}-\eqref{eq:curl-s_i}, the vector potential is not sourced by any of these components alone but is correlated with the rotational part of the full momentum density field. This panel also shows that the distribution of the vector potential magnitude is a great deal smoother than the cases of matter and velocity fields. This is expected since the vector potential components satisfy the elliptic-type equation~\eqref{eq:MDC-CF}, and then long-wavelength modes become dominant due to the Laplacian operator $\partial^2$. Although not included here, the same happens in the case {of the conformal factor $\psi$ which satisfies the Hamiltonian constraint (or the Poisson equation in the Newtonian limit)}. From the quantitative side, we note that the vector potential magnitude seems to typically remain between $\mathcal{O}(10^{-8})$ and $\mathcal{O}(10^{-7})$, with some peaks of a few times $\mathcal{O}(10^{-7})$ only in very specific regions.

We will explore the behaviour of the vector modes in more detail in the next sections.

\subsection{Power spectra}
\label{sec:spectra}

In this section we analyse the power spectra of the velocity field and gravitomagnetic vector potential. The auto and cross spectra of matter quantities such as density, velocity divergence and vorticity (which are measured with \dtfe\ from particle data) are calculated using {\sc nbodykit} \citep{nbodykit:2018}. In contrast, the vector (as well as scalar) potential values are calculated and stored by \gramses\ in cells of hierarchical AMR meshes, and the spectrum is measured by a different code that is able to handle such mesh data directly and to write it on a regular grid by interpolation. 
While the vector potential spectrum can also be measured in the same way as the matter quantities by writing its values at the particles' positions rather than in AMR cells, which means \dtfe~ and {\sc nbodykit} can be used, the above method yields better results on small scales as shown in Appendix~\ref{appendix}.
In all figures, we normalise the velocity power spectra by the factor $(\mathcal{H}f)^2$, where $\mathcal{H}=aH$ is the conformal Hubble parameter of the reference Friedmann universe, and $f$ the linear growth rate in $\Lambda$CDM parameterised as~\citep{Linder-growth}
\begin{equation}
f(a)=\Omega_m(a)^{6/11},
\end{equation}
where $\Omega_m(a)=\Omega_{m}a^{-3}/(H/H_0)^2$. In this way, the amplitude of $P_{\theta}$ matches that of the matter power spectrum in the linear regime, where the continuity equation $\delta=-\theta/(\mathcal{H}f)$ is expected to hold. 


\begin{figure*}
\centering
\includegraphics[width=\linewidth]{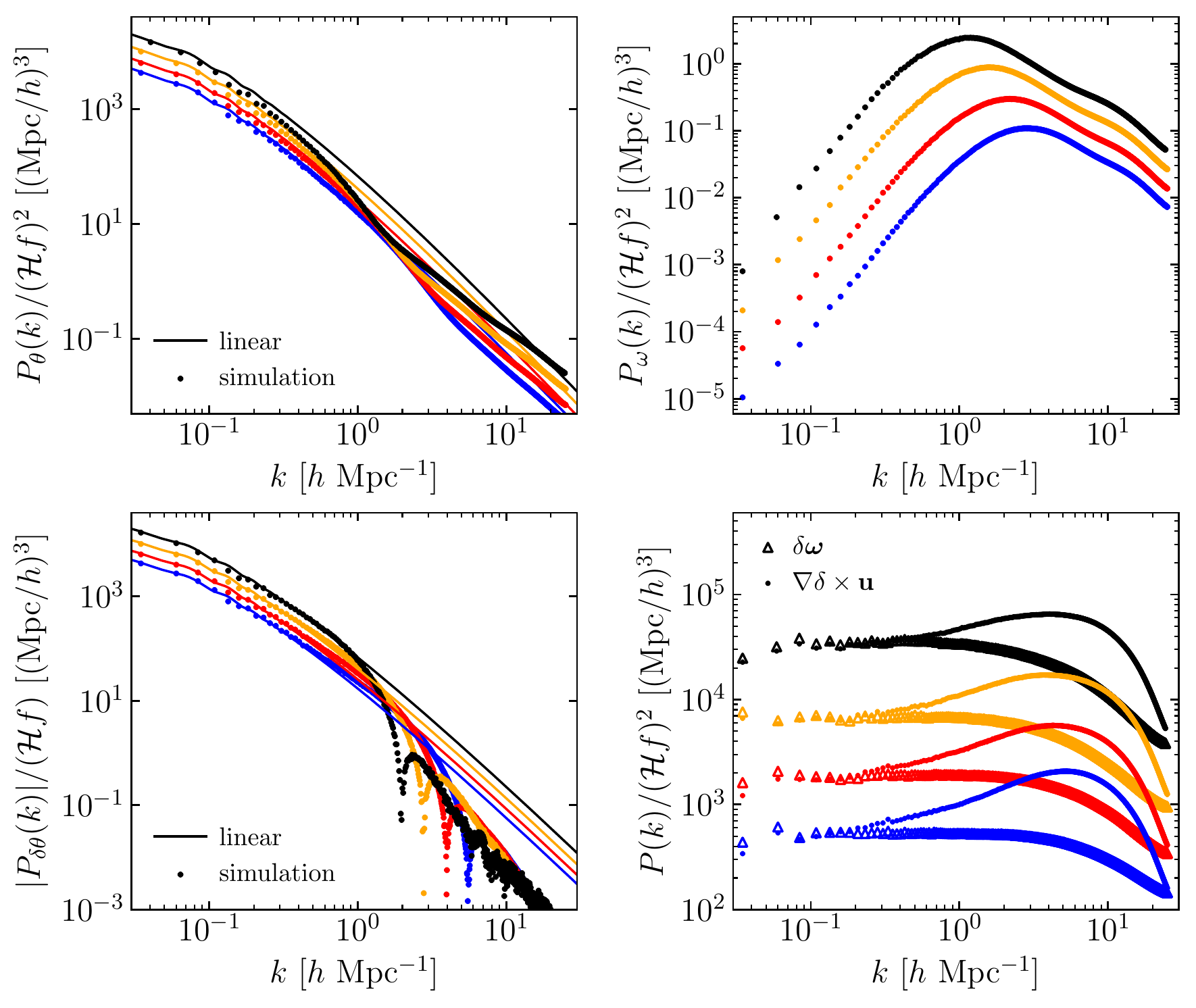}
\caption{(Colour Online) Various auto and cross power spectra involving the velocity field for $z=0$ (black), $z=0.5$ (orange), $z=1$ (red) and $z=1.5$ (blue). The top left and top right panels show the velocity divergence power spectrum and vorticity power spectrum, respectively, both of which are normalised by $(\mathcal{H}f)^2$. Bottom left: the cross spectrum between density and velocity divergence. Since in linear theory $P_{\delta\theta}<0$, we plot its absolute magnitude normalised by $\mathcal{H}f$. The discontinuity corresponds to the flip in sign on nonlinear scales, after which density and velocity divergence become correlated. Bottom right: the power spectrum of $\delta\boldsymbol{\omega}$ and $\boldsymbol{\nabla}\delta\times{\bf u}$, which are the main source terms for the metric vector potential, c.f.~Eq.~\eqref{eq:curl-s_i}. These are normalised by $(\mathcal{H}f)^2$. In the two left panels, the solid lines denote the corresponding linear-theory predictions.}
\label{fig:vel-spectra}
\end{figure*}

Figure~\ref{fig:vel-spectra} shows the velocity divergence power spectrum (top left panel), the vorticity power spectrum (top right panel), the cross spectrum between density and velocity divergence (bottom left) and the power spectrum of two different contributions to the momentum field (bottom right) at different redshifts in the range $0\leq z\leq1.5$. In the case of velocity divergence, we find a very good agreement with linear theory at scales $k\leq 0.1\invMpch$ for all redshifts. Above that scale, deviations become stronger towards lower redshifts, and a localised power loss (`dip') eventually develops around $k\approx1.2\invMpch$. 
In the case of the vorticity power spectrum, we note that towards large scales this is several orders of magnitude smaller than velocity divergence, while at around $k\sim1\invMpch$ 
{the spectrum} starts to peak and they become comparable. 
Note that, unlike the velocity divergence, there is no standard perturbation theory prediction for the vorticity as this exactly vanishes in the perfect fluid description. Interestingly, the `dip' in the velocity divergence power spectrum is at the similar position to the peak in the vorticity power spectrum, which has been interpreted as the consequence of shell crossing occurring around that scales, where the angular momentum can be large enough to dampen the growth of structures as it forces particles to rotate around them~\citep{jelic-cizmek_generation_2018}. 

Note that, due to the high cost\footnote{A GR simulation using {\gramses} takes about an order of magnitude longer than an equivalent Newtonian simulation using default {\sc ramses}, partly due to the 10 (compared to one) GR metric potentials to be solved, and partly due to the cost of preparing the source terms for the nonlinear equations that govern the  metric potentials, as well as the additional {\sc mpi} communications.} of GR simulations using \gramses, we have not performed runs with even higher resolutions to check the convergence of the velocity and vorticity power spectra. A useful convergence test for {\it gevolution} simulations was done in \citet{jelic-cizmek_generation_2018} (see Fig.~6 there), which shows that the amplitude of $P_\omega$ decreases as the {force resolution increases}. The simulations there have the same box size of $L_{\rm box}=256\Mpch$, and the run labelled `high resolution 1' has the same mesh resolution as our domain grid ($1024^3$ cells); while this resolution is eight times poorer than that of the run labelled `high resolution 2', which has $2048^3$ cells, the AMR nature of \gramses\ means that higher resolution can be achieved in high-density regions -- with the highest resolution attained in our run being equivalent to a regular mesh with $128^3\times1024^3$ cells. Hence, since `high resolution' 1 and 2 are already converged in \citet{jelic-cizmek_generation_2018}, we conclude that our simulation has also converged to at least a similar level.

The cross spectra $P_{\delta\theta}$ is useful for detecting deviations from linear theory and provides information about shell crossing. Considering the continuity equation, the linear-theory expectation is that $P_{\theta\delta}/(\mathcal{H}f)=-P_{\delta}$, but towards shell-crossing scales the initial (linear) anti-correlation of $\delta$ and $\theta$ is lost and correlations appear~\citep{hahn_properties_2015}. From the bottom left panel of Fig.~\ref{fig:vel-spectra} we find that the anti-correlation drops dramatically and flips sign at $k\approx2\invMpch$ at $z=0$, which is slightly higher than the scale at which the vorticity spectrum peaks as also found in previous studies~\citep{jelic-cizmek_generation_2018}.

The bottom right panel of Fig.~\ref{fig:vel-spectra} shows the power spectra of $\delta\boldsymbol{\omega}$ and $\boldsymbol{\nabla}\delta\times{\bf u}$, which are the main source terms for the metric vector potential in Eq.~\eqref{eq:curl-s_i}. In particular, the contribution of $\boldsymbol{\omega}$ to Eq.~\eqref{eq:curl-s_i} is already small compared to $\delta\boldsymbol{\omega}$ on nonlinear scales because $\delta\gg1$. We find good agreement with the $z=0$ results shown in~\citet{bruni_computing_2014} based on a post-Friedmann expansion. We find that towards higher redshifts the contribution due to $\boldsymbol{\nabla}\delta\times{\bf u}$ starts to become larger than that of $\delta\boldsymbol{\omega}$ at slightly larger scales.

Although vorticity vanishes in standard perturbation theory, the effective field theory of LSS (EFTofLSS) predicts that its power spectrum today can be characterised by a power law over a range of scales~\citep{Carrasco:2013mua}. On large scales, we can find the slope of the vorticity power spectrum by fitting a power law,
\begin{equation}
    P_{\boldsymbol{\omega}}(k)=A_{\omega}k^{n_{\omega}}\,,\label{eq:Pk_vort-large-scales}
\end{equation}
where $n_\omega$ is the large-scale spectral index, and $A_\omega$ the amplitude that is not fixed by theory. The EFTofLSS predicts $n_{\omega}=3.6$ for $0.1\invMpch\lesssim k\lesssim0.2\invMpch$ and  $n_{\omega}=2.8$ for $0.2\invMpch\lesssim k\lesssim0.6\invMpch$ \citep{Carrasco:2013mua}. Previous $N$-body simulations have found $n_{\omega}\approx2.5$ for $k\lesssim0.1\invMpch$ \citep{hahn_properties_2015}; a similar value was found at $k\lesssim0.4\invMpch$ in \citet{jelic-cizmek_generation_2018}. Moreover, on scales $k\gtrsim 1\invMpch$, there is partial evidence suggesting that the spectral index approaches the asymptotic value $n^{\rm NL}_{\omega}\to-1.5$ \citep{hahn_properties_2015}. 

\begin{figure}
\centering
\includegraphics[width=\linewidth]{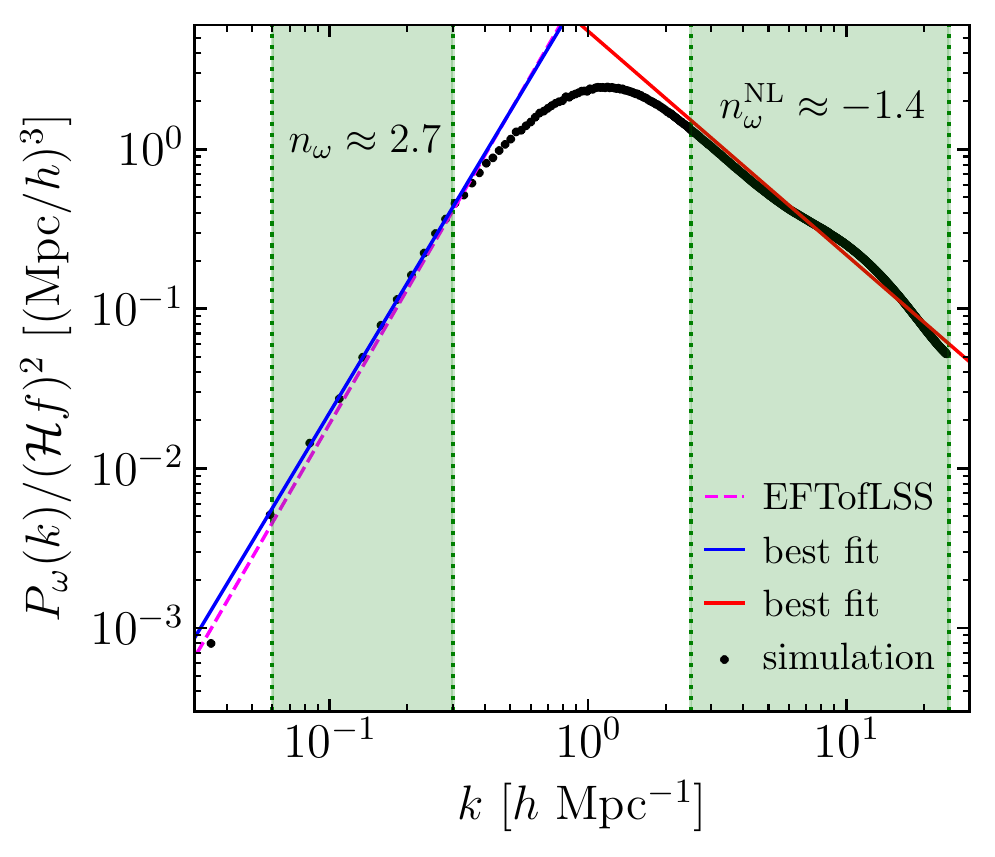}
    \caption{(Colour Online) Power-law 
    {fitting} of the vorticity power spectrum at $z=0$. The solid blue and solid red lines show the best fits of the simulation data (black dots) on large and small scales, respectively, while the shaded regions represent the validity interval for each fit. As a reference, the dashed magenta line shows the EFTofLSS prediction from~\citet{Carrasco:2013mua} for the region $0.2\invMpch\protect\lesssim k\protect\lesssim0.6\invMpch$, which only has a small overlap with the fitting region used on large sub-horizon scales.}
    \label{fig:vorticity-power-law-fit}
\end{figure}

Figure~\ref{fig:vorticity-power-law-fit} shows the best fits of the power law \eqref{eq:Pk_vort-large-scales} to the simulation data at $z=0$ on large scales (small scales) with their corresponding spectral index $n_\omega$ ($n^{\rm NL}_{\omega}$), and the shaded region represents the interval of validity for the fit. On large sub-horizon scales, we find $n_\omega\approx2.7$, which is slightly higher than previous simulations results in the literature, and slightly lower than the EFTofLSS prediction. Notice, however, that there is not complete overlap between the region used for the fit and the EFTofLSS prediction used for comparison as the latter extends up to $k\sim0.6\invMpch$ but it is clear that the slope of the power spectrum already decreases at 
$k\sim0.32\invMpch$. In addition, the slope does not seem to become steeper at larger scales as predicted by the EFTofLSS, a feature also found by the previous study \citep{jelic-cizmek_generation_2018}, which is likely related to the large-scale cutoff imposed by the finite box of the simulation. 
Toward smaller scales, we find the spectral index $n^{\rm NL}_\omega\approx-1.4$, which is slightly less steep than that suggested in \citet{hahn_properties_2015}. However, there is a slight but clear increase in power at around $k\sim7\invMpch$ which introduces an oscillatory feature not captured by a perfect power law.

\begin{figure*}
\centering
\includegraphics[width=\linewidth]{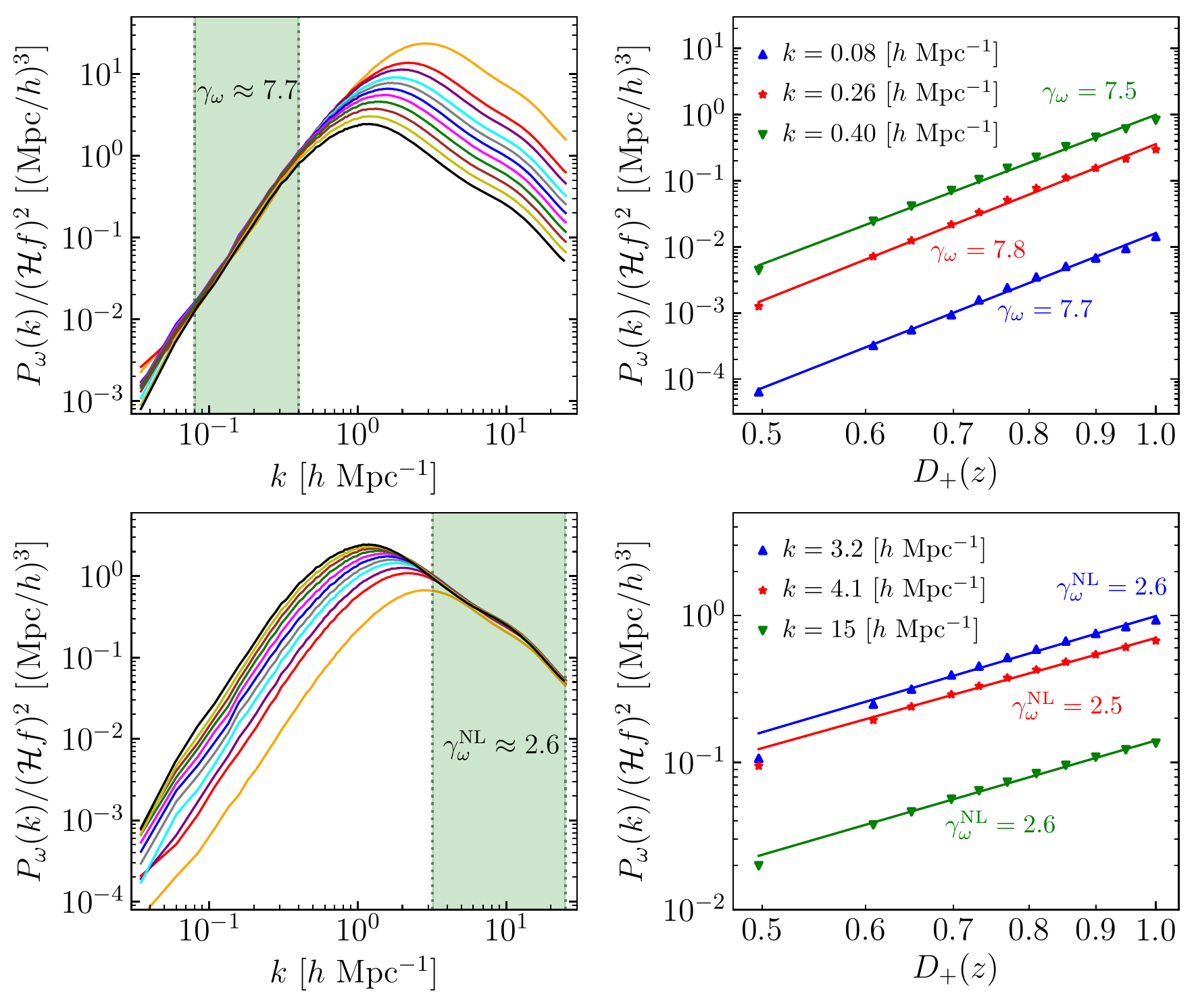}
    \caption{(Colour Online) Power-law modelling of the time evolution of the vorticity power spectrum based on Eq.~\eqref{eq:Pk_vort_evol}. Top panels show results for the large scales regime and the bottom panels analogous results for nonlinear scales. Top left: vorticity power spectra at different redshifts scaled using $\gamma_\omega=7.7$. Shaded regions represent the interval of validity considered for the fit, and the colors $\{$orange, red, purple, cyan, gray, blue, magenta, green, brown, yellow, brown, black$\}$ correspond to $z=\{$1.5, 1, 0.85, 0.7, 0.6, 0.5, 0.4, 0.3, 0.2, 0.1, 0$\}$, respectively. Bottom left: similar to top left panel but for nonlinear scales. Right panels: Time evolution of the vorticity power spectrum for a set of fixed $k$-modes as a function of $D_+(z)$ (normalised by today's value of $D_+$). The solid lines correspond the best fit curves with the respective power-law indices $\gamma_\omega$ and $\gamma^{\rm NL}_\omega$ shown. On the bottom right panel, the data point for $z=1.5$ has not been included for the fit, as the bottom left panel shows a clear discrepancy with lower redshifts.}
    \label{fig:vorticity-power-law-fit-NL}
\end{figure*}

As originally proposed in~\citet{pueblas_generation_2009}, it is also interesting to characterise the evolution of the large-scale vorticity power spectrum as 
\begin{equation}
 P_{\boldsymbol{\omega}}(k;z)= \left(\frac{D_{+}(z)}{D_{+}(0)}\right)^{\gamma_\omega}P_{\boldsymbol{\omega}}(k;z=0)\,,\label{eq:Pk_vort_evol}
\end{equation}
where $D_{+}(z)$ is the linear growth rate at $z$ and $\gamma_\omega$ a new parameter. In~\citet{pueblas_generation_2009}, the best-fit value found is $\gamma_\omega=7\pm0.3$ using the snapshots $z=0,1,3$, which is overall consistent with \citet{thomas_fully_2015,jelic-cizmek_generation_2018}, although the latter references suggest values $\gamma_\omega\geq7$. Moreover, these have only considered snapshots with $z\leq1$ since the scaling breaks down at higher 
redshifts, which is likely related to resolution effects in the sampling of vorticity due to a lower fraction of particles undergoing shell crossing at higher redshifts. 

The top panels of Fig.~\ref{fig:vorticity-power-law-fit-NL} show the results for the best fist of the $D^{\gamma_\omega}_+$ scaling of Eq.~\eqref{eq:Pk_vort_evol} using several snapshots below $z=1.5$. The top left panel of Fig.~\ref{fig:vorticity-power-law-fit-NL} shows the power spectrum at these various redshifts scaled using $(D_{+}(z)/D_{+}(0))^{7.7}$, while in the top right panel we select three different modes from the shaded green region of the top left panel and find the corresponding value of $\gamma_\omega$ from a best fit to the corresponding vorticity spectra.
We find that there is some scale dependence in $\gamma_\omega$ and the amplitude of the vorticity power spectrum evolves approximately with $\gamma\approx{7.7}$ over the scales $0.08\lesssim k \lesssim 0.4$, which is higher than other simulation results in the literature~\citep{pueblas_generation_2009,thomas_fully_2015,jelic-cizmek_generation_2018}. However, compared to the latter two references, in the case here we are able to fit the amplitude up to $z=1.5$ before the scaling breaks down. Besides the results from~\citet{jelic-cizmek_generation_2018} based on the {\it gevolution} code, which works in a fixed-resolution grid, previous studies of vorticity use $N$-body simulation codes in which a softening length scale in the force calculation determines the spatial resolution. In the case of \gramses, the AMR {capabilities} allow one to achieve high spatial resolution ($\sim 2\kpch$) in high-density regions.

We can extend the previous analysis to model the time evolution of the vorticity power spectrum at nonlinear scales, in terms of a new scale-independent parameter $\gamma^{\rm NL}_\omega$ in Eq.~\eqref{eq:Pk_vort_evol}. From Fig.~\ref{fig:vel-spectra}, it is clear that the power spectrum evolves more slowly in this regime compared with large scales, and so we expect $\gamma^{\rm NL}_\omega$ to be smaller than $\gamma_\omega$. In the bottom left panel of Fig.~\ref{fig:vorticity-power-law-fit-NL}, we show the scaling of the vorticity spectra by $(D_{+}(z)/D_{+}(0))^{2.6}$, where we find that such evolution works as a good approximation on scales $k\gtrsim3.2\invMpch$. In the bottom right panel we show the best-fit value of $\gamma^{\rm NL}_\omega$ for three different $k$-modes. In this case, unlike in the previous fit for large sub-horizon scales, we have not considered the $z=1.5$ spectrum for the fit as from the bottom left panel it is already clear that the scaling for such spectrum (orange solid line) would deviate from the lower redshift results.
This result suggests that the amplitude of the vorticity power spectrum can be actually estimated using a scale-independent parameter in the power law of Eq.~\eqref{eq:Pk_vort_evol} on deeply nonlinear scales. However, there is an obvious scale dependence in the transition between the large- and small-scale regimes which is not captured by 
these parameterisations and requires further investigation. 


\begin{figure*}
    \centering
    \includegraphics[width=\linewidth]{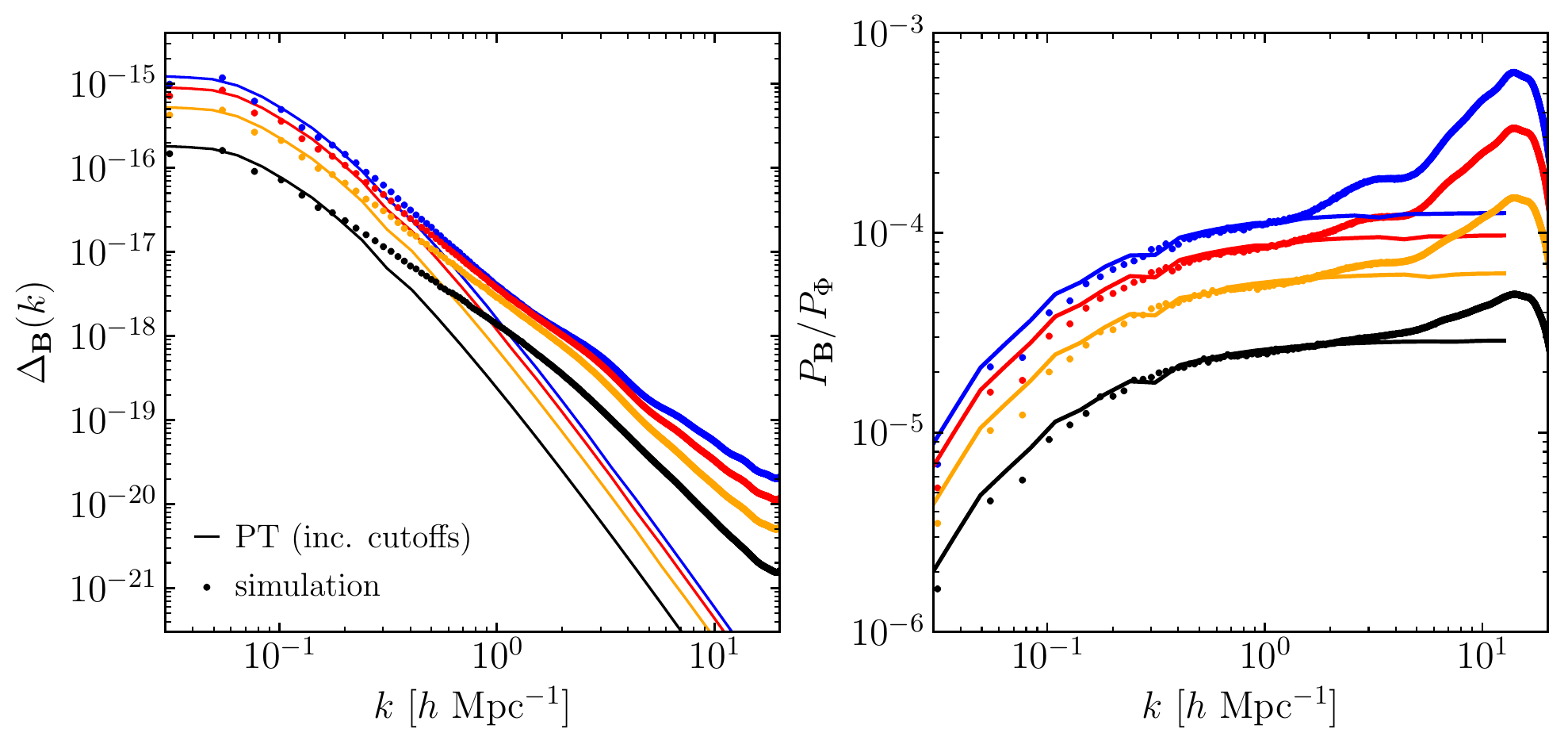}
    \caption{(Colour Online) Left: The dimensionless power spectrum of the vector potential, $\Delta_{\bf B}(k)=k^3P_{\bf B}(k)/(2\pi^2)$. The solid lines represent the corresponding second-order perturbation theory predictions~\citep{Lu:2008ju}, in which cutoffs have been introduced in the convolution calculation to accommodate the lack of power in the simulation results on large scales due to box size. 
    Right: The ratio between the power spectrum of the vector potential and that of the scalar gravitational potential defined as the fully nonlinear perturbation to the lapse function, i.e., $\Phi\equiv\alpha-1$. 
    In both panels, each colour corresponds to $z=0$ (black), $z=0.5$ (orange), $z=1$ (red) and $z=1.5$ (blue).} 
    \label{fig:shift-vector-power}
\end{figure*}

Let us now discuss the results for the vector potential. In $\Lambda$CDM cosmology, this appears as a second-order perturbation at its lowest order, which in the case of a perfect fluid is sourced by the product of the first-order density contrast and velocity divergence \citep{Matarrese:1997ay,Lu:2008ju}. However, the single-stream fluid description of CDM breaks down at late times when shell crossing occurs, and then we expect corrections to the vector potential particularly at quasi-linear and nonlinear scales.

The second-order perturbation theory prediction for the dimensionless power spectrum of ${\bf B}$,
\begin{equation}
    \Delta_{\bf B}(k) \equiv \frac{k^3}{2\pi^2}P_{\bf B}(k),
\end{equation} 
is given by~\citep{Lu:2008ju}
\begin{align}
\label{eq:delta-beta-v-convolution}
\Delta_{{\bf B}}(k)=& \frac{9\Omega^2_mH^4_0}{2a^2k^2}\int^\infty_0{\rm d}w\,\times\\
&\int^{1+w}_{|1-w|}{\rm d}u\Pi\left[\Delta_{\delta}(ku)\Delta_{v}(kw)-\frac{w}{u}\Delta_{\delta v}(ku)\Delta_{\delta v}(kw)\right]\,,\nonumber
\end{align}
where $\Delta_\delta$ and $\Delta_v$ are the dimensionless power spectra of the density perturbation and velocity potential $v$, $\Delta_{\delta v}$ their cross spectrum, and $\Pi(u,w)=u^{-2}w^{-4}\left[4w^2-(1+w^2-u^2)^2\right]$ is an integration kernel that depends on $w=k'/k$ and  $u=\sqrt{1+w^2-2w\cos{\vartheta}}$, with $\cos\vartheta$ defined by 
{$\cos{\vartheta}={\bf k'}\cdot{\bf k}/(kk')$}. 
At any given scale, the convolution in Eq.~\eqref{eq:delta-beta-v-convolution} couples different $k$-modes of $\delta$ and $v$.
%
Since the simulation can only access modes within a
finite $k$-range, this is equivalent to having a large-scale ($k_{\rm min}$) and small-scale ($k_{\rm max}$) cutoffs in Eq.~\eqref{eq:delta-beta-v-convolution}, therefore leading to a lower amplitude of $P_{\bf B}$ than the true result. For instance, \citet{gevolution-main} found that in order to get good agreement between simulation results and perturbation-theory calculations using Eq.~\eqref{eq:delta-beta-v-convolution}, the box should be large enough to contain the matter-radiation equality scale. In practice, to account for this effect due to missing $k$-modes, to compare with Eq.~\eqref{eq:delta-beta-v-convolution},
%
we use the large-scale cutoff $k_{\rm min}\sim{0.8\times2\pi/L}$, i.e. 80 percent of the fundamental mode of the box, as well as a small-scale cutoff $k_{\rm max}=\pi N_{\rm part}^{1/3}/L$, which corresponds to the Nyquist wavenumber of the coarsest grid used by the simulation. 
The left panel of Fig.~\ref{fig:shift-vector-power} shows the simulation measurements of the dimensionless power spectrum of the vector potential at four different redshifts, and their corresponding perturbation-theory predictions.
At 
{$z\geq1$} we see good agreement between the simulation and perturbation-theory results up to 
{$k\sim0.3\invMpch$}, while at $z=0$ discrepancies start already at $k\sim0.2\invMpch$, which is qualitatively consistent with \citet{Adamek:2014xba,bruni_computing_2014}; see also~\citet{Andrianomena:2014} for a prescription of the nonlinear corrections to the perturbation-theory result using {\sc halofit}. At highly nonlinear scales the amplitude of the spectrum measured from the simulation can be 
{more than} two orders of magnitude 
{higher} than the perturbation-theory prediction. Note that at all four redshifts the simulation spectra flatten at the largest $k$-mode sampled by the simulation box, which can be interpreted as a finite-box effect. 

The right panel of Fig.~\ref{fig:shift-vector-power} shows the ratio between the power spectra of vector potential ${\bf B}$ and that of the scalar potential $\Phi$ measured from the simulation, the latter defined as the fully nonlinear perturbation to the lapse function in the metric \eqref{eq:adm-metric}, i.e. $\Phi\equiv\alpha-1$. At $z=0$, we find the ratio to be within $2\times10^{-5}$ and $4\times10^{-5}$ 
for $0.2\invMpch\lesssim k\lesssim10\invMpch$, which is in good agreement with~\citet{bruni_computing_2014}. The ratio reaches a peak of $5\times10^{-5}$ at $k\sim15\invMpch$, after which it starts to decrease. At higher redshift the evolution of ${\bf B}$ makes the ratio larger. Our results confirm that the ratio between both potentials reach the percent-level on nonlinear scales {at $z=0$}. As pointed out by~\citet{bruni_computing_2014}, though this ratio is close to the value found in~\citet{Lu:2008ju} for the ratio between scalar and vector modes in perturbation theory, here the fully nonlinear ${\bf B},\Phi$ fields are used. In fact, the vector potential power spectrum from the left panel of Fig.~\ref{fig:shift-vector-power} can be over two orders of magnitude larger than that found in the latter reference.

\subsection{The vector potential and frame-dragging acceleration in dark matter haloes}
\label{sec:haloes}

Let us further analyse the vector potential on nonlinear scales by investigating its magnitude inside the dark matter haloes from the above general-relativistic simulation. For this we have generated halo catalogues using the phase-space Friends-of-Friends halo finder {\sc rockstar}~\citep{rockstar-2013}. From this catalogue we then get their centre positions, radii $R_{200c}$ and masses $M_{200c}$. The latter two are defined respectively as the distance from the halo centre which encloses a mean density of 200 times the critical density of the universe as a given redshift, and the mass enclosed within such a sphere.

Unfortunately, the {inaccuracy} when estimating the velocity divergence and vorticity fields on small scales using \dtfe~prevents us from studying their behaviour in haloes alongside the vector potential. We have tested that indeed, the velocity estimations are strongly affected by resolution and do not converge either using a resolution for the tessellation grid similar to the 
mean inter-particle distance of dark matter particles in the haloes or otherwise. The phase-interpolation method was used in~\citet{hahn_properties_2015} to successfully estimate the vorticity in haloes in the case of warm dark matter, but still it is not possible to robustly measure this from CDM simulations either: this is related to the difficulty of resolving the perturbations up to highly nonlinear scales in the CDM case, which in warm dark matter models is not required as the spectrum truncates at some finite free-streaming scale.

\begin{figure*}
\includegraphics[width=\linewidth]{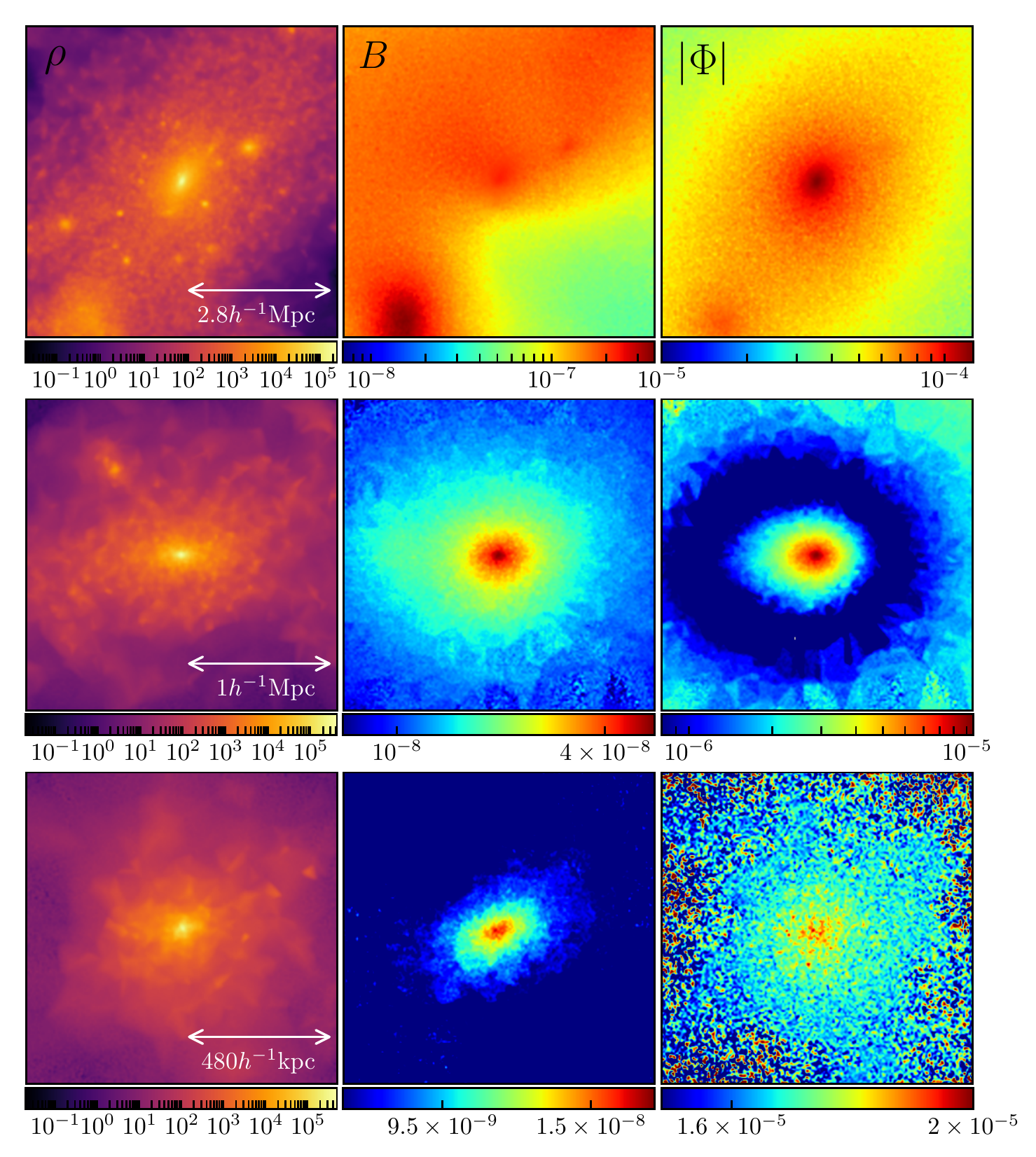}
\caption{(Colour Online) Visualisation of three selected dark matter haloes at $z=0$, with masses $M_h=6.5\times10^{14}\Msolar$ (top row), $M_h=3.0\times10^{13}\Msolar$ (middle row) and $M_h=3.1\times10^{12}\Msolar$ (bottom row). In each row, each panel shows, from left to right: matter density, magnitude of the vector potential and absolute magnitude the scalar gravitational potential (since typically $\Phi\leq0$ in the inner parts of a halo). Interpolation has been used to display smoother maps. All maps are in logarithmic scale.}
\label{fig:halo-maps}
\end{figure*}

\begin{figure*}
\includegraphics[width=\linewidth]{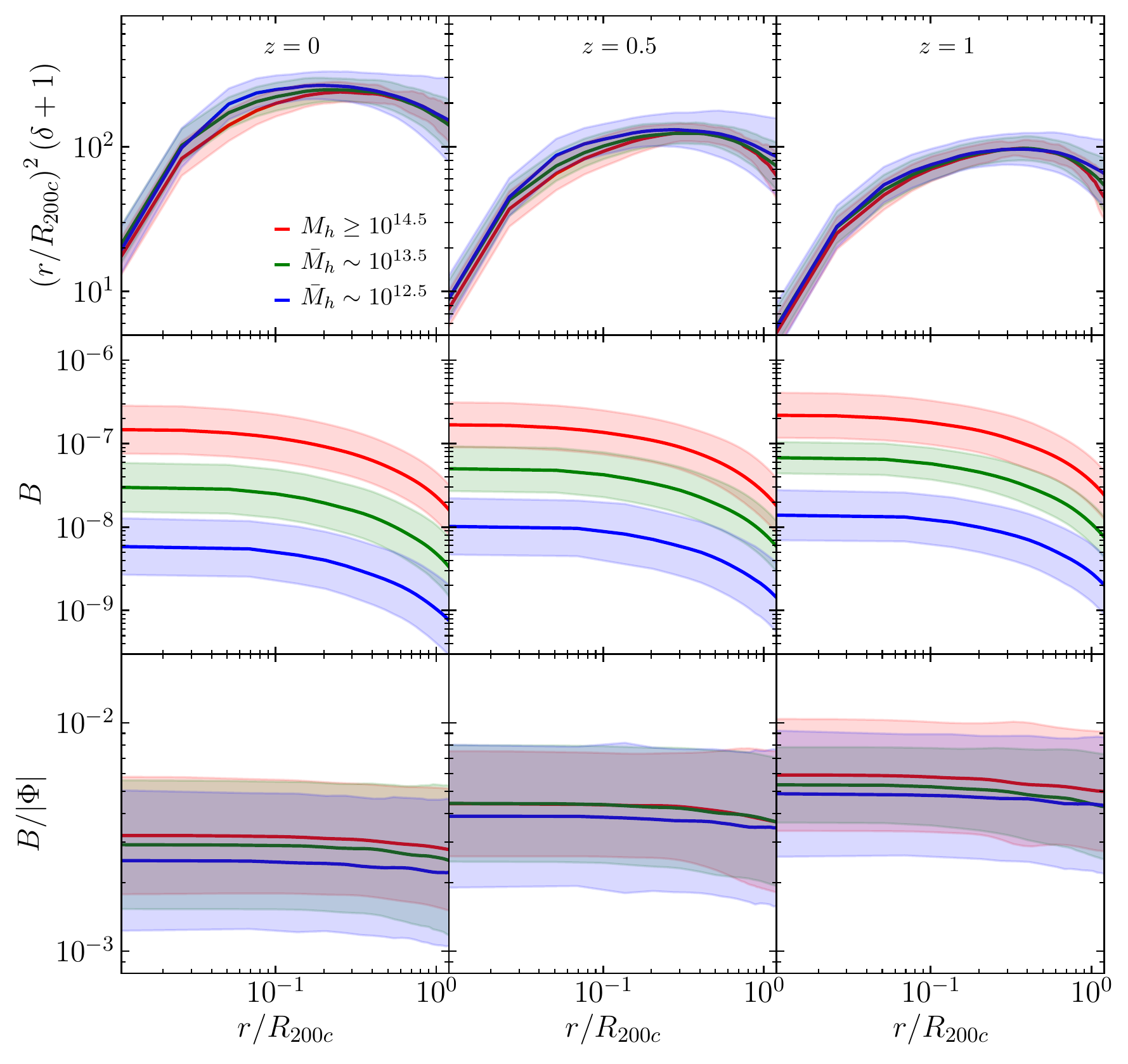}
\caption{(Colour Online) Halo profiles (spherical averages) at $z=0$ (left column), $z=0.5$ (middle column) and $z=1$ (right column). Each row shows, from top to bottom, density, vector potential magnitude and its ratio against the scalar gravitational potential.
In the case of the potentials, their spherical-average at $R_{200c}$ has been subtracted from each individual halo profile as a way to remove their environmental contributions. The upper, middle and lower halo mass ranges are represented by red, green and blue, respectively, for which the solid line shows the mean calculated over all haloes in a given mass range, and the shaded regions are the $1\sigma$ regions. The values of $M_h$ shown in the inset are in units of$\Msolar$.}
\label{fig:halo-profiles}
\end{figure*}

Figure~\ref{fig:halo-maps} shows density (left column), vector potential magnitude (middle column) and scalar gravitational potential (right column) in the vicinity of three selected dark matter haloes at $z=0$, with masses  $M_h\approx6.5\times10^{15}\Msolar$ (top row), $M_h\approx 3.0\times10^{13}\Msolar$ (middle row) and $M_h\approx 3.1\times10^{12}\Msolar$ (bottom row). In all cases, the map centre is aligned with the halo centre and the width of the shown region corresponds to four times the halo radius $R_{200c}$.
As also shown in Fig~\ref{fig:maps}, overall we observe some degree of correlation between the vector potential and the matter density, but 
clearly not at the level of the scalar potential.
In particular, in the case of the most massive halo (top row) we can see that while both potentials peak towards the halo centre, unlike for the scalar potential, the global maximum of the vector potential within the shown region is actually found 
in the lower left part of the map, where there appears to be another, smaller, halo infalling towards the central one. Again, this qualitative difference is not surprising since the vector potential is sourced by the transverse part of the momentum density, Eq.~\eqref{eq:curl-s_i}, while the matter source term for the scalar potential is the density contrast itself (up to higher-order terms). As before, we can also see that both potentials are smoother than the density field owing to the elliptic-type nature of their 
equations~\citep{gramses-1}, in which short-wavelength modes are 
dominated. In addition, in the most massive halo we can observe that the scalar potential tends to be more spherically symmetric around the center than $B$, which displays large values in most part of the left and upper part of the map. 
Indeed, although the low-density (dark) regions in the bottom right and top left parts of the density map are of similar characteristics, and these are clearly correlated with the $\Phi$ map, these are not correlated with features in the $B$ map at all. 

For the halo shown in the middle 
row of Fig.~\ref{fig:halo-maps}, the density and potential contours have more similar shapes to each other than in the most massive halo. Nonetheless, the scalar potential again seems to decay more rapidly outside $R_{200c}$ than the vector potential magnitude. This also seems to be the case in the halo shown in the bottom panels, although in this case the potentials are smaller and shallower. Note that, for the halo in the middle panels, $|\Phi|$ is largest in the central region (red/orange/green), decays when one moves further away from the halo centre (blue), but grows again far from the halo (green); this is because this halo resides in a low-density environment, with a positive environmental contribution to the total potential so that the latter crosses zero.

It is important to bear in mind that, although the halo centres are approximately located at a local maximum of $\vert\Phi\vert$, the potentials themselves are not an observable quantity: it is the gradient of the potentials that contributes as force terms in the geodesic equation \eqref{eq:geodesic-eq-1}, while the values of the potential themselves 
can be largely influenced by their environments.
In this subsection, we are mainly interested in haloes which are isolated and therefore less affected by environments. To select such haloes, we try to split the potential at each point into two contributions: one from the halo itself and one from its environment, i.e., well beyond a distance $R_{200c}$ from its centre. Since the potentials are not necessarily spherically symmetric, as it is evident from the top row of Fig.~\ref{fig:halo-maps}, as a crude way, we shall take the spherical average in a radial bin at $2R_{200c}$ and subtract this from the values at smaller radii, which allows to get ``shifted'' radial halo profiles for both $\Phi$ and $B$ that vanish at $2R_{200c}$. For $\Phi$ ($B$) we expect this profile to monotonically increase (decrease) to zero as $r$ increases to $2R_{200c}$, for well-isolated relaxed haloes. 

\begin{figure}
\centering
\includegraphics[width=\linewidth]{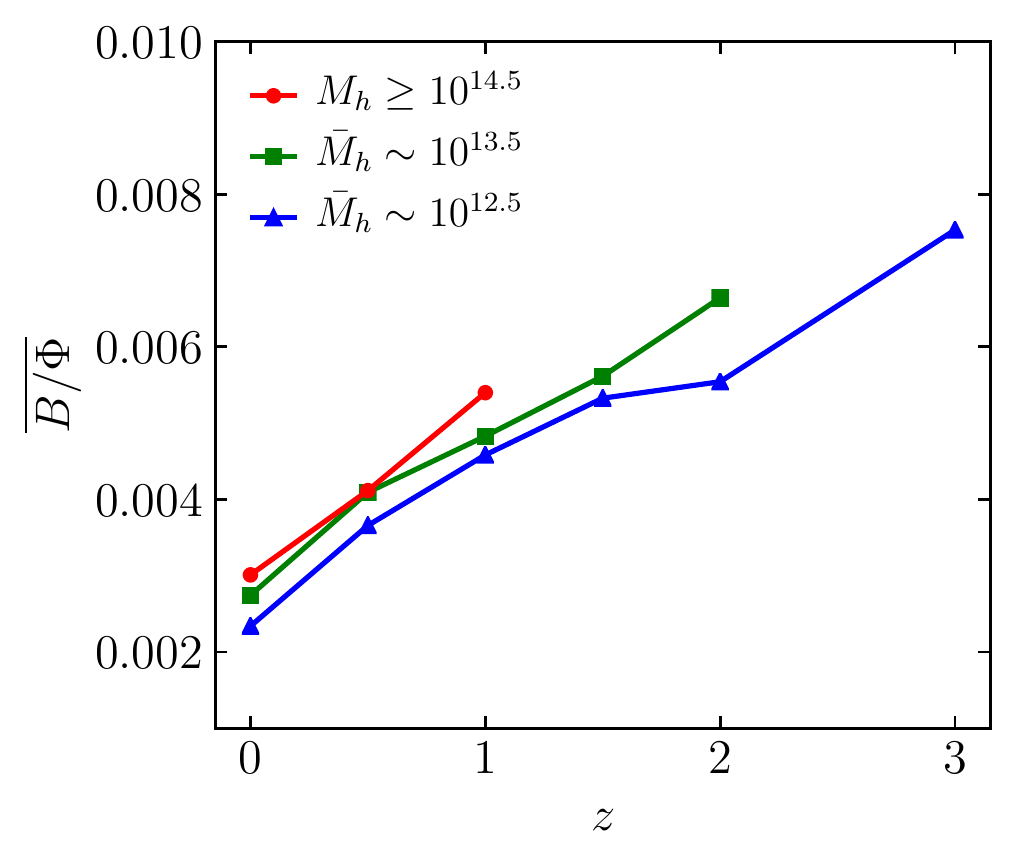}
    \caption{(Colour Online) Evolution of the ratio between the vector potential and the scalar gravitational potential for the different halo mass ranges. At each redshift, the value shown corresponds to the average of the ratio for $r\leq R_{200c}$. We have only included cases where the number of haloes in a given mass range is greater than ten at a given redshift. The values of $M_h$ shown in the inset are in units of$\Msolar$.}
    \label{fig:B_over_Phi_evolution}
\end{figure}

\begin{figure*}
\includegraphics[width=\linewidth]{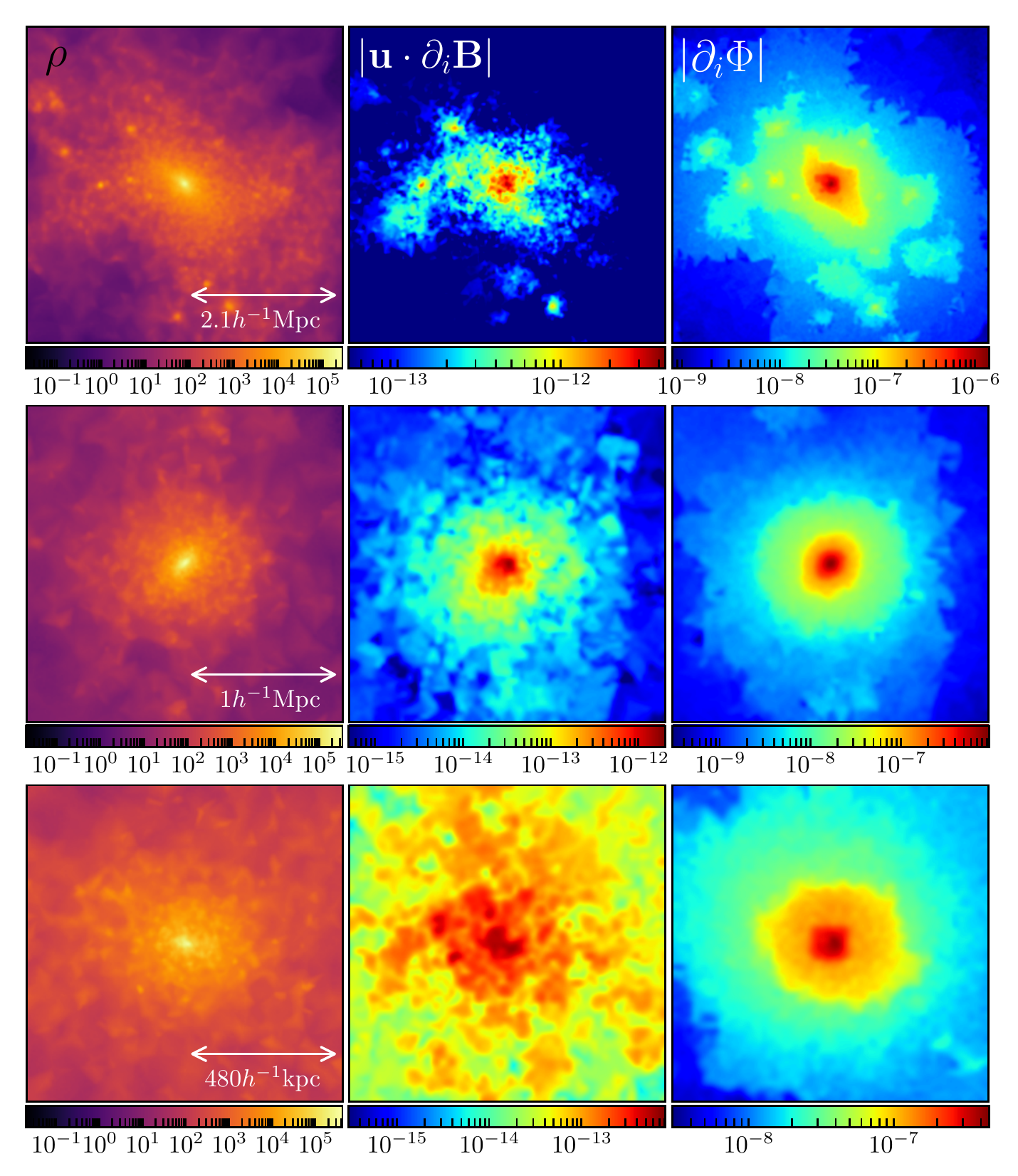}
\caption{(Colour Online) Visualisation of three selected dark matter haloes at $z=0$, with masses $M_h=2.7\times10^{14}\Msolar$ (top row), $M_h=3.3\times10^{13}\Msolar$ (middle row) and $M_h=3.2\times10^{12}\Msolar$ (bottom row). In each row, each column shows, from left to right: matter density, the magnitude of the gravitomagnetic acceleration and the magnitude of the standard gravitational acceleration, the latter two in units of $h$ cm$/$s$^2$. Interpolation has been used to display smoother maps. All maps are in logarithmic scale.}
\label{fig:halo-maps-accel}
\end{figure*}

\begin{figure*}
\includegraphics[width=\linewidth]{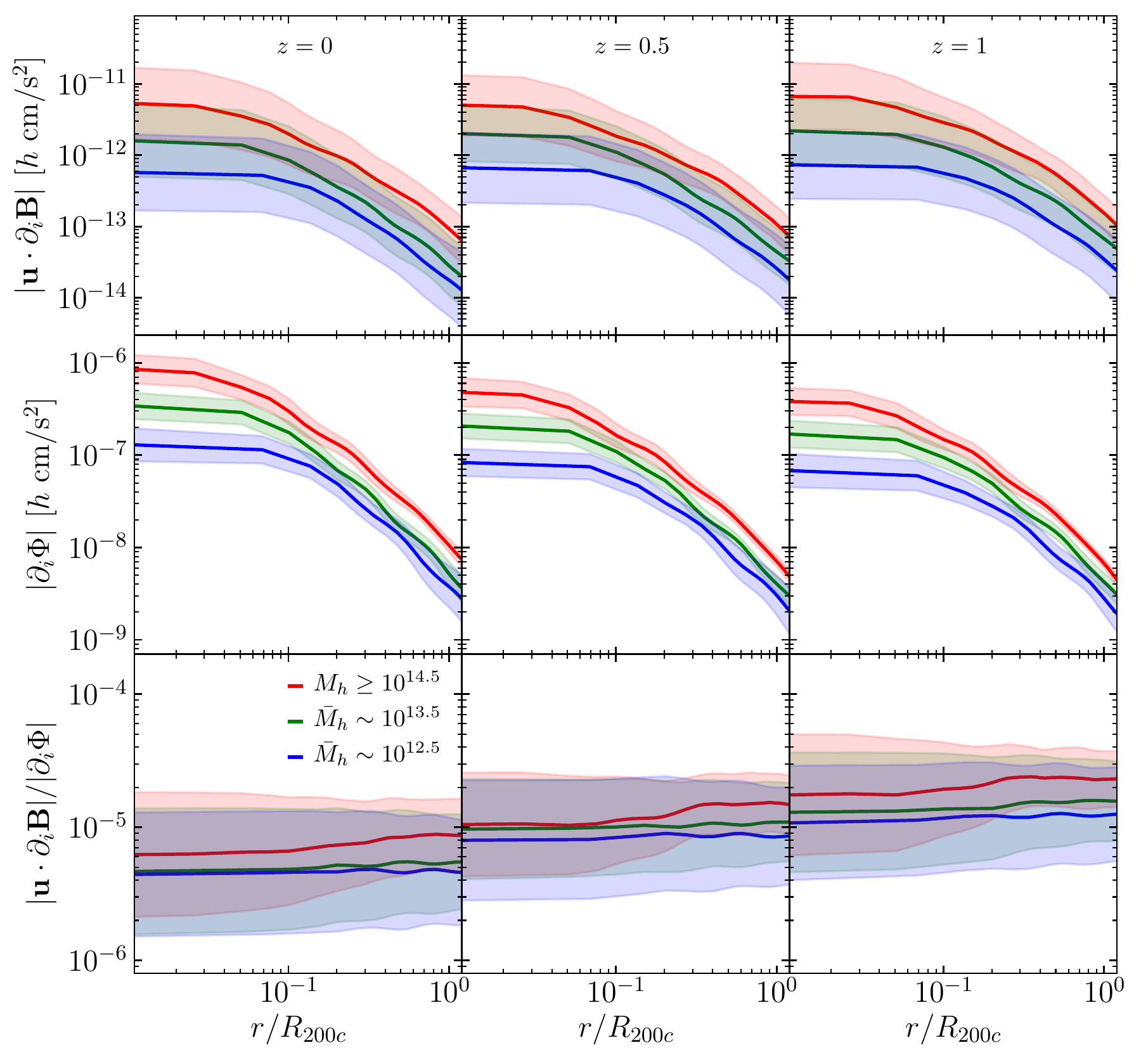}
\caption{(Colour Online) Halo profiles (spherical averages) at $z=0$ (left column), $z=0.5$ (middle column) and $z=1$ (right column). In a given column, each row shows, from top to bottom, the gravitomagnetic (frame-dragging) acceleration, standard gravitational acceleration and their ratio. The upper, middle and lower halo mass ranges are represented by red, green and blue, respectively, for which the solid line shows the mean calculated over all haloes in a given mass range, and the shaded regions are the $1\sigma$ regions. The values of $M_h$ shown in the inset are in units of$\Msolar$.}
\label{fig:halo-profiles-accel}
\end{figure*}

Figure~\ref{fig:halo-profiles} shows, from the top to the bottom row, the radial profiles of density, the vector potential magnitude and its ratio against the scalar gravitational potential. All 
profiles have been measured from the centres of a sample of haloes in different mass ranges, for three redshifts: $z=0$ (left column), $z=0.5$ (middle column) and $z=1$ (right column). For this we have 
selected three subsamples of haloes with $\mathcal{O}(100)$ haloes each based on mass cuts: we define a higher mass range $M_h\geq10^{14.5}\Msolar$, an intermediate mass range with mean mass $\bar{M}_h=10^{13.5}\Msolar$, and a lower mass range with mean mass $\bar{M}_h=10^{12.5}\Msolar$. For each halo from a given mass range, we then calculate the spherical average of the density, vector potential and scalar potential up to $2R_{200c}$, and average over the full population.  
As mentioned in the previous paragraph, in the case of the potentials we have subtracted their average values at $2R_{200c}$ in the profile of each individual halo. 
In this process, we have discarded the haloes 
in which the resulting spherical average of 
$B$ becomes negative for some $r<R_{200c}$ after the subtraction, which typically happens in lower mass haloes due to their shallow potentials.
%
However, these haloes are the most abundant type and hence we retain a sample of size $\mathcal{O}(100)$ even at $z=1$, while the number of haloes in the middle and higher mass bins is around $\sim50$ at that same redshift.

From Fig.~\ref{fig:halo-profiles} we find that at the $1\sigma$ level there is a clear correlation between halo mass and the magnitude of the gravitomagnetic potential, which can differ by up to two orders of magnitude between halos with masses close to $10^{12.5}\Msolar$ and those with masses larger than $10^{14.5}\Msolar$. In all cases, the vector potential flattens toward the halo centres and it decreases towards the outskirts. However, from the bottom row of Fig.~\ref{fig:halo-profiles} we find that the ratio between vector and scalar potentials is roughly constant inside haloes across all masses and redshifts considered, and the dependence of this ratio upon halo mass is quite weak as all means lie within $1\sigma$ of each other. At $z=0$, we find that the ratio is a few times $10^{-3}$, which is roughly consistent with the value inferred from the ratio of $\mathcal{O}(10^{-5})$ between the power spectra of the vector and scalar potentials at $k\gtrsim\mathcal{O}(0.1)\invMpch$, as shown in Fig.~\ref{fig:shift-vector-power} (note that the subtraction of the environmental contributions in these potentials essentially removes the long-wavelength contributions to $\overline{B/\Phi}$, thereby marking this comparison with Fig.~\ref{fig:shift-vector-power} reasonable; but as we only look at a small fraction of the total volume, inside a sub-group of haloes, we of course should not expect an exact equality). 
At $z=0.5$ and $z=1$, the picture is qualitatively the same apart from the increase in the amplitude of the vector potential.

In CDM simulations, it is well known that the density profile of haloes can be described by the universal Navarro-Frenk-White \citep[NFW;][]{Navarro:1995iw} fitting formula, which has a corresponding analytical prediction for the Newtonian potential profiles of haloes. The constancy of $B/|\Phi|$ inside haloes which is found here implies that it might be straightforward to derive an analytical fitting function for the $B$ profiles in haloes, which is closely related to the NFW function, though this will not be pursued in this paper.

Given that Fig.~\ref{fig:halo-profiles} shows that the ratio between the vector and scalar potentials is roughly constant inside the halos -- and we have checked that such constant ratio holds even above $z=1$ -- we can characterise this ratio by a single number at each halo mass and redshift. As an extension of the bottom row of Fig.~\ref{fig:halo-profiles}, Fig.~\ref{fig:B_over_Phi_evolution} shows the mean value of such ratio calculated within $r<R_{200c}$ at different redshifts. Since the number of haloes in a given mass bin decreases towards higher redshifts, here we only consider cases in which the number of haloes in a given mass range is greater than ten at a given redshift. We find that for all mass bins $\overline{B/\Phi}$ increases almost linearly with redshift. At redshift $z=2$ the 
rate of change of this ratio with respect to redshift slows down slightly for the lowest mass range (blue line), after which it picks up again: this could be due to a lack of simulation resolution at high $z$.
{Observationally, the ratio between vector and scalar potentials is particularly relevant for weak lensing, as post-Newtonian calculations show that the relative correction to the Newtonian convergence field $\kappa$ is proportional to $B/\Phi$~\citep{Sereno:2002,Sereno:2003,bruni_computing_2014}. Therefore, Fig.~\ref{fig:B_over_Phi_evolution} suggests that, in the case of dark matter haloes, the lensing convergence correction due to the gravitomagnetic potential is between the $\mathcal{O}(10^{-3})$ and $\mathcal{O}(10^{-2})$ level, in agreement with previous studies~\citep{Sereno:2007,Cuesta-Lazaro:2018,Tang_gravitomagnetic:2020}. Moreover, this only depends weakly on the halo mass and could be more easily detected on high-mass haloes at high redshifts. 
However, we note that at higher orders in the post-Newtonian expansion, new contributions from the time derivative of $B$ appear~\citep{bruni_computing_2014,Thomas:2014aga} as well, which requires further inspection.}

Besides investigating the potentials, we can also look at the force that each of these exert on the particles according to Eq.~\eqref{eq:geodesic-eq-1}, which shows that the total force is mainly composed by two contributions; the standard gravitational force arising from the gradient of the scalar potential (first term on the r.h.s.), and the gravitomagnetic force (contained in the second term on the r.h.s.) which is responsible for the frame-dragging effect. The latter is naturally not taken into account
in Newtonian gravity. The third term in the r.h.s of Eq.~\eqref{eq:geodesic-eq-1} is subdominant and so we shall not explore it here.

Figure~\ref{fig:halo-maps-accel} is a visualisation of the magnitude of the gravitomagnetic acceleration (middle column) and that of the standard gravitational acceleration (right column) in units of $h$ cm$/$s$^2$, in the vicinity of three different dark matter haloes. These haloes have similar masses to those shown in Fig.~\ref{fig:halo-maps}. We find that the forces are correlated with the density field up to some degree, particularly in the haloes in the middle and bottom rows, although the gravitomagnetic force seems to be less smooth than the Newtonian one.
For the halo in the top row, there is a clearer difference between the forces compared to the other two cases. The peaks of the gravitomagnetic acceleration seem to occur at the density peaks but the opposite is not true, and there is no clear correspondence between their amplitudes. 
Interestingly, in this halo the values of gravitomagnetic force around a few times $10^{-13}$ $h$ cm$/$s$^2$ (green region) extend around the centre and towards the left part of the map, where the density field has already decreased by various orders of magnitude. This kind of asymmetry between both kinds of maps might be due to the actual dynamical state of the particles in a given region. Even if the density is low, if the particles' velocity happens to be aligned with the gradient of the vector potential components they will contribute significantly to $\vert{\bf u}\cdot\partial_i{\bf B}\vert$. 

As before, we can calculate the spherical averages of the forces, which allows us to get radial profiles (although no subtraction from radial bins beyond $2R_{200c}$ is required this time). Figure~\ref{fig:halo-profiles-accel} shows a comparison of the gravitomagnetic (frame-dragging) acceleration and the standard gravitational one in dark matter haloes in an analogous way to the scalar and vector potential profiles shown in Fig.~\ref{fig:halo-profiles}.
We find that the magnitude of the gravitomagnetic force is larger towards the inner parts of the halo, and the dependence on the halo mass is weaker than in the case of the scalar gravitational potential. As we discussed before, this can also be explained by the fact that the gravitomagnetic force not only depends on density but on the actual dynamical state of particles. 
Similarly to the behaviour of $B/|\Phi|$, from Fig.~\ref{fig:halo-profiles-accel} we find that the ratio of the two corresponding forces also remains fairly constant inside the haloes, although in the most massive haloes it tends to increase toward the outskirts. A weak dependence on halo mass is found at all redshifts.
In~\citet{gevolution-main} the maximum gravitomagnetic acceleration measured from the simulation box at $z=0$ is found to be roughly $7\times10^{-12}$ $h$ {cm/s$^2$} for the highest resolution used ($125\kpch$), while the value measured from lower resolution runs decreases monotonically. From Fig.~\ref{fig:halo-profiles-accel} we find that this is comparable with our results for haloes in the upper mass range at the $1\sigma$ level. However, we note that for the most massive halo in our simulation, we find the maximum value of the gravitomagnetic acceleration to be $7\times10^{-11}$ $h$ {cm/s$^2$}, i.e. roughly one order of magnitude higher. This difference could be explained by the fact that in our simulation the most refined regions are resolved with a resolution of $2\kpch$. In addition, \gramses~treats the vector potential non-perturbatively, although the difference due to higher-order corrections is likely to be subdominant with respect to the aforementioned resolution dependence.

\section{Conclusions}

We have investigated the vector modes of the matter fields as well as those of the $\Lambda$CDM spacetime metric, from large sub-horizon scales to deeply nonlinear scales using a high-resolution run of the general-relativistic $N$-body \gramses~code~\citep{gramses-1,gramses-2}. On the one hand, vorticity vanishes at the non-perturbative level in a perfect fluid description and yet it is generated dynamically due to the collisionless nature of dark matter. On the other hand, the metric vector potential -- responsible for frame-dragging -- appears beyond linear order in perturbation theory and is not solved for in Newtonian simulations. Therefore, the physics behind the vector modes is highly non-trivial and numerical simulations play an important role in their study. Although the relativistic nature of the code is not particularly exploited from the point of view of vorticity, the vector potential is a prime quantity as this is not part of Newtonian gravity and therefore not implemented in Newtonian simulations. 

To this end, we have run a high-resolution $N$-body simulation using \gramses, that employs $N_{\rm part}=1024^3$ particles in a box of comoving size $L_{\rm box}=256\Mpch$. In \gramses, the GR metric potentials -- in the fully constrained formalism and conformally flat approximation -- are solved on meshes in configuration space. The AMR capabilities of \gramses\ allows it to start off with a regular grid with $1024^3$ cells, and hierarchically refine it in high-density regions to reach a spatial resolution of $2\kpch$ in the most refined places, namely dark matter haloes. This enables a quantitative analysis of the behaviour of vector modes in such regions.

The key findings of this paper are summarised as follows:
\begin{enumerate}
    \item On scales $0.06\invMpch\lesssim k\lesssim0.3\invMpch$, the vorticity power spectrum can be characterised by the power law in Eq.~\eqref{eq:Pk_vort-large-scales} with an index $n_\omega\approx2.7$, a value that is overall consistent with 
    recent simulation results of \citet{hahn_properties_2015,jelic-cizmek_generation_2018}. 
    On nonlinear scales ($2.3\invMpch\lesssim{k}\lesssim20\invMpch$), the power spectrum can again be described by a power-law function, but the index changes to $n^{\rm NL}_\omega\approx-1.4$, close to the asymptotic value of $-1.5$ suggested by \citet{hahn_properties_2015}; cf.~Fig.~\ref{fig:vorticity-power-law-fit}.
    \item On scales $0.1\invMpch\lesssim k\lesssim0.4\invMpch$ the amplitude of the vorticity power spectrum seems to evolve as $\sim[D_+(z)/D_+(0)]^{7.7}$ at $z\leq1.5$, which is higher than previous values found in the literature~\citep{thomas_fully_2015,jelic-cizmek_generation_2018}. Nonetheless, these references also found larger values than the scaling with the seventh power originally proposed in \citet{pueblas_generation_2009}. On scales $k\gtrsim3.5\invMpch$, the evolution of the amplitude of the power spectrum can be similarly neatly described as $\sim[D_+(z)/D_+(0)]^{2.6}$
    up to $z=1$; cf.~Fig.~\ref{fig:vorticity-power-law-fit-NL}.
    \item The vector potential power spectrum remains below $4\times10^{-5}$ relative to the scalar gravitational potential down to $k=20\invMpch$; cf.~Fig.~\ref{fig:shift-vector-power}.
    \item Inside dark matter haloes, the magnitude
    of the vector potential peaks towards the centres at $\sim10^{-7}$ for haloes more massive than $10^{14.5}\Msolar$, which can reduce by two orders of magnitude in haloes of masses around $10^{12.5}\Msolar$. Its ratio against the scalar gravitational potential remains typically a few times $10^{-3}$ inside the haloes, regardless of their mass (cf.~Fig.~\ref{fig:halo-profiles}). {The ratio $B/|\Phi|$ remains nearly flat within the halo radius $R_{200c}$, for the halo redshift ($z<3$) and mass $(10^{12.5}\sim10^{15}\Msolar)$ ranges checked, and this constant increases roughly linearly with $z$; cf.~Fig.~\ref{fig:B_over_Phi_evolution}.}
    \item The magnitude of the gravitomagnetic acceleration also peaks at the halo centres where it can reach a few times $10^{-11}$ $h$ cm$/$s$^2$ in haloes above $\sim10^{14.5}\Msolar$. 
    Its ratio against the standard gravitational acceleration remains around $\sim10^{-5}$ on average, regardless of the halo mass and distance from the halo centre; {cf.~Fig.~\ref{fig:halo-profiles-accel}.}
    This suggests that the effect of the gravitomagnetic force on cosmic structure formation is, even for the most massive structures, negligible -- however, note that we have not studied the behaviour in low-density regions, i.e., voids. 
    \end{enumerate}

While we have presented a first study of the gravitomagnetic potential in dark matter haloes with general-relativistic simulations, there are several possible extensions in this direction. The analysis of the gravitomagnetic potential and forces done in this paper could be extended to galaxies, e.g., by constructing a catalogue using certain 
semi-analytic models. {It is then possible to calculate the gravitomagnetic accelerations of galaxies based on their coordinates and velocities. However, as we have seen above, this acceleration is much weaker than the standard gravitational acceleration, and} the impact of baryons on small scales still remains to be assessed. The 
{implementation} of 
(magneto)hydrodynamics in the default {\sc ramses} code could be used in conjunction with the general-relativistic implementation of \gramses~as a first approximation {to address this question, although we generally expect that uncertainties in baryonic physics should surpass GR effects.}

A {perhaps} more interesting possibility is 
to self-consistently implement massive neutrinos and radiation in this relativistic code. In the second \gramses\ code paper \citep{gramses-2}, we have introduced a method to generate initial conditions for \gramses\ simulations that does not require back-scaling. It is therefore natural to evolve these matter components which are neglected in traditional simulations \citep[e.g.,][]{Adamek:2017uiq}. On the same vein, a Newtonian (quasi-static) implementation of modified gravity models on \gramses~would allow to study the gravitomagnetic potential in such type of theory. In particular, the modified gravity code {\sc ecosmog}~\citep{Li2012:ECOSMOG,Li2013:ECOSMOG} is based on {\sc ramses} and can be easily made compatible with \gramses\ for such purpose.

In this paper, we have primarily focused on the general-relativistic physical quantities that could impact cosmic structure formation, and this can ultimately 
only be observed by detecting photons~\citep{McDonald:2009,Croft:2013MNRAS,Bonvin:2014-asymm,Shadab:2017MNRAS}. Therefore, 
besides the gravitomagnetic force acting on massive particles, it is also important to study how vector modes, {as well as other GR effects, could influence} the photon trajectories on nonlinear scales, and what is the consequent impact on observables, e.g.\ lensing \citep{Thomas:2014aga,Saga:2015apa,Gressel:2019jxw}. This requires the implementation of general-relativistic ray tracing algorithms \citep[e.g.][]{Barreira:2016wqo,Breton:2018wzk,Lepori:2020ifz,reverdy-thesis} and is left as a future project.

\section*{Acknowledgements}

We thank Marius Cautun for assistance with the \dtfe\ code, and Ra\'ul Angulo for useful discussions on the vorticity estimation from $N$-body simulations. We are also grateful to James Mertens and to the anonymous referee for their valuable comments and observations.

CB-H is supported by the Chilean National Agency of Research and Development (ANID) through grant CONICYT/Becas-Chile (No.~72180214). BL is supported by the European Research Council (ERC) through ERC starting Grant No.~716532, and STFC Consolidated Grant (Nos.~ST/I00162X/1, ST/P000541/1). MB  is  supported  by  UK  STFC Consolidated Grant No.  ST/S000550/1.

This work used the DiRAC@Durham facility managed by the Institute for Computational Cosmology on behalf of the STFC DiRAC HPC Facility (\url{www.dirac.ac.uk}). The equipment was funded by BEIS via STFC capital grants ST/K00042X/1, ST/P002293/1, ST/R002371/1 and ST/S002502/1, Durham University and STFC operation grant ST/R000832/1. DiRAC is part of the UK National e-Infrastructure.

\section*{Data Availability}

For access to the simulation data please contact CB-H.



\bibliographystyle{mnras}
\bibliography{gramses_B_modes} 

\begin{thebibliography}{}
\makeatletter
\relax
\def\mn@urlcharsother{\let\do\@makeother \do\$\do\&\do\#\do\^\do\_\do\%\do\~}
\def\mn@doi{\begingroup\mn@urlcharsother \@ifnextchar [ {\mn@doi@}
  {\mn@doi@[]}}
\def\mn@doi@[#1]#2{\def\@tempa{#1}\ifx\@tempa\@empty \href
  {http://dx.doi.org/#2} {doi:#2}\else \href {http://dx.doi.org/#2} {#1}\fi
  \endgroup}
\def\mn@eprint#1#2{\mn@eprint@#1:#2::\@nil}
\def\mn@eprint@arXiv#1{\href {http://arxiv.org/abs/#1} {{\tt arXiv:#1}}}
\def\mn@eprint@dblp#1{\href {http://dblp.uni-trier.de/rec/bibtex/#1.xml}
  {dblp:#1}}
\def\mn@eprint@#1:#2:#3:#4\@nil{\def\@tempa {#1}\def\@tempb {#2}\def\@tempc
  {#3}\ifx \@tempc \@empty \let \@tempc \@tempb \let \@tempb \@tempa \fi \ifx
  \@tempb \@empty \def\@tempb {arXiv}\fi \@ifundefined
  {mn@eprint@\@tempb}{\@tempb:\@tempc}{\expandafter \expandafter \csname
  mn@eprint@\@tempb\endcsname \expandafter{\@tempc}}}

\bibitem[\protect\citeauthoryear{Abel, Hahn  \& Kaehler}{Abel
  et~al.}{2012}]{Abel_2012}
Abel T.,  Hahn O.,   Kaehler R.,  2012, \mn@doi [Monthly Notices of the Royal
  Astronomical Society] {10.1111/j.1365-2966.2012.21754.x}, 427, 61–76

\bibitem[\protect\citeauthoryear{Adamek, Durrer  \& Kunz}{Adamek
  et~al.}{2014}]{Adamek:2014xba}
Adamek J.,  Durrer R.,   Kunz M.,  2014, \mn@doi [Class. Quant. Grav.]
  {10.1088/0264-9381/31/23/234006}, 31, 234006

\bibitem[\protect\citeauthoryear{Adamek, Daverio, Durrer  \& Kunz}{Adamek
  et~al.}{2016a}]{Adamek:2016zes}
Adamek J.,  Daverio D.,  Durrer R.,   Kunz M.,  2016a, \mn@doi [JCAP]
  {10.1088/1475-7516/2016/07/053}, 07, 053

\bibitem[\protect\citeauthoryear{{Adamek}, {Daverio}, {Durrer}  \&
  {Kunz}}{{Adamek} et~al.}{2016b}]{gevolution-main}
{Adamek} J.,  {Daverio} D.,  {Durrer} R.,   {Kunz} M.,  2016b, \mn@doi [Nature
  Physics] {10.1038/nphys3673}, \href
  {https://ui.adsabs.harvard.edu/abs/2016NatPh..12..346A} {12, 346}

\bibitem[\protect\citeauthoryear{Adamek, Durrer  \& Kunz}{Adamek
  et~al.}{2017}]{Adamek:2017uiq}
Adamek J.,  Durrer R.,   Kunz M.,  2017, \mn@doi [JCAP]
  {10.1088/1475-7516/2017/11/004}, 11, 004

\bibitem[\protect\citeauthoryear{{Adamek}, {Barrera-Hinojosa}, {Bruni}, {Li},
  {Macpherson}  \& {Mertens}}{{Adamek} et~al.}{2020}]{Adamek:2020jmr}
{Adamek} J.,  {Barrera-Hinojosa} C.,  {Bruni} M.,  {Li} B.,  {Macpherson}
  H.~J.,   {Mertens} J.~B.,  2020, \mn@doi [Classical and Quantum Gravity]
  {10.1088/1361-6382/ab939b}, \href
  {https://ui.adsabs.harvard.edu/abs/2020CQGra..37o4001A} {37, 154001}

\bibitem[\protect\citeauthoryear{{Alam}, {Zhu}, {Croft}, {Ho}, {Giusarma}  \&
  {Schneider}}{{Alam} et~al.}{2017}]{Shadab:2017MNRAS}
{Alam} S.,  {Zhu} H.,  {Croft} R. A.~C.,  {Ho} S.,  {Giusarma} E.,
  {Schneider} D.~P.,  2017, \mn@doi [Monthly Notices of the Royal Astronomical
  Society] {10.1093/mnras/stx1421}, \href
  {https://ui.adsabs.harvard.edu/abs/2017MNRAS.470.2822A} {470, 2822}

\bibitem[\protect\citeauthoryear{{Andrianomena}, {Clarkson}, {Patel}, {Umeh}
  \& {Uzan}}{{Andrianomena} et~al.}{2014}]{Andrianomena:2014}
{Andrianomena} S.,  {Clarkson} C.,  {Patel} P.,  {Umeh} O.,   {Uzan} J.-P.,
  2014, \mn@doi [\jcap] {10.1088/1475-7516/2014/06/023}, \href
  {https://ui.adsabs.harvard.edu/abs/2014JCAP...06..023A} {2014, 023}

\bibitem[\protect\citeauthoryear{Bardeen}{Bardeen}{1980}]{Bardeen:1980kt}
Bardeen J.~M.,  1980, \mn@doi [Phys. Rev. D] {10.1103/PhysRevD.22.1882}, 22,
  1882

\bibitem[\protect\citeauthoryear{Barreira, Llinares, Bose  \& Li}{Barreira
  et~al.}{2016}]{Barreira:2016wqo}
Barreira A.,  Llinares C.,  Bose S.,   Li B.,  2016, \mn@doi [JCAP]
  {10.1088/1475-7516/2016/05/001}, 05, 001

\bibitem[\protect\citeauthoryear{{Barrera-Hinojosa} \& {Li}}{{Barrera-Hinojosa}
  \& {Li}}{2020a}]{gramses-1}
{Barrera-Hinojosa} C.,  {Li} B.,  2020a, \mn@doi [JCAP]
  {10.1088/1475-7516/2020/01/007}, \href
  {https://ui.adsabs.harvard.edu/abs/2020JCAP...01..007B} {2020, 007}

\bibitem[\protect\citeauthoryear{{Barrera-Hinojosa} \& {Li}}{{Barrera-Hinojosa}
  \& {Li}}{2020b}]{gramses-2}
{Barrera-Hinojosa} C.,  {Li} B.,  2020b, \mn@doi [JCAP]
  {10.1088/1475-7516/2020/04/056}, \href
  {https://ui.adsabs.harvard.edu/abs/2020JCAP...04..056B} {2020, 056}

\bibitem[\protect\citeauthoryear{{Behroozi}, {Wechsler}  \& {Wu}}{{Behroozi}
  et~al.}{2013}]{rockstar-2013}
{Behroozi} P.~S.,  {Wechsler} R.~H.,   {Wu} H.-Y.,  2013, \mn@doi [\apj]
  {10.1088/0004-637X/762/2/109}, \href
  {https://ui.adsabs.harvard.edu/abs/2013ApJ...762..109B} {762, 109}

\bibitem[\protect\citeauthoryear{Bertschinger}{Bertschinger}{1993}]{Bertschinger:1993xt}
Bertschinger E.,  1993, in {Les Houches Summer School on Cosmology and Large
  Scale Structure (Session 60)}. pp 273--348 (\mn@eprint {arXiv}
  {astro-ph/9503125})

\bibitem[\protect\citeauthoryear{Bonazzola, Gourgoulhon, Grandcl\'ement  \&
  Novak}{Bonazzola et~al.}{2004}]{Bonazzola-FCF:2004}
Bonazzola S.,  Gourgoulhon E.,  Grandcl\'ement P.,   Novak J.,  2004, \mn@doi
  [Phys. Rev. D] {10.1103/PhysRevD.70.104007}, 70, 104007

\bibitem[\protect\citeauthoryear{Bonvin, Hui  \& Gazta\~naga}{Bonvin
  et~al.}{2014}]{Bonvin:2014-asymm}
Bonvin C.,  Hui L.,   Gazta\~naga E.,  2014, \mn@doi [Phys. Rev. D]
  {10.1103/PhysRevD.89.083535}, 89, 083535

\bibitem[\protect\citeauthoryear{Bonvin, Durrer, Khosravi, Kunz  \&
  Sawicki}{Bonvin et~al.}{2018}]{bonvin_redshift-space_2018}
Bonvin C.,  Durrer R.,  Khosravi N.,  Kunz M.,   Sawicki I.,  2018, \mn@doi
  [Journal of Cosmology and Astroparticle Physics]
  {10.1088/1475-7516/2018/02/028}, 2018, 028

\bibitem[\protect\citeauthoryear{{Breton}, {Rasera}, {Taruya}, {Lacombe}  \&
  {Saga}}{{Breton} et~al.}{2019}]{Breton:2018wzk}
{Breton} M.-A.,  {Rasera} Y.,  {Taruya} A.,  {Lacombe} O.,   {Saga} S.,  2019,
  \mn@doi [\mnras] {10.1093/mnras/sty3206}, \href
  {https://ui.adsabs.harvard.edu/abs/2019MNRAS.483.2671B} {483, 2671}

\bibitem[\protect\citeauthoryear{Bruni, Thomas  \& Wands}{Bruni
  et~al.}{2014}]{bruni_computing_2014}
Bruni M.,  Thomas D.~B.,   Wands D.,  2014, \mn@doi [Physical Review D]
  {10.1103/PhysRevD.89.044010}, 89, 044010

\bibitem[\protect\citeauthoryear{Carrasco, Foreman, Green  \&
  Senatore}{Carrasco et~al.}{2014}]{Carrasco:2013mua}
Carrasco J. J.~M.,  Foreman S.,  Green D.,   Senatore L.,  2014, \mn@doi [JCAP]
  {10.1088/1475-7516/2014/07/057}, 07, 057

\bibitem[\protect\citeauthoryear{{Cautun} \& {van de Weygaert}}{{Cautun} \&
  {van de Weygaert}}{2011}]{Cautun:2011-DTFE}
{Cautun} M.~C.,  {van de Weygaert} R.,  2011, arXiv e-prints, \href
  {https://ui.adsabs.harvard.edu/abs/2011arXiv1105.0370C} {p. arXiv:1105.0370}

\bibitem[\protect\citeauthoryear{Cordero-Carri\'on, Cerd\'a-Dur\'an,
  Dimmelmeier, Jaramillo, Novak  \& Gourgoulhon}{Cordero-Carri\'on
  et~al.}{2009}]{CorderoCarrion:2008nf}
Cordero-Carri\'on I.,  Cerd\'a-Dur\'an P.,  Dimmelmeier H.,  Jaramillo J.~L.,
  Novak J.,   Gourgoulhon E.,  2009, \mn@doi [Phys. Rev. D]
  {10.1103/PhysRevD.79.024017}, 79, 024017

\bibitem[\protect\citeauthoryear{Crocce, Pueblas  \& Scoccimarro}{Crocce
  et~al.}{2006}]{Crocce2006:2LPT}
Crocce M.,  Pueblas S.,   Scoccimarro R.,  2006, \mn@doi [Mon. Not. Roy.
  Astron. Soc.] {10.1111/j.1365-2966.2006.11040.x}, 373, 369

\bibitem[\protect\citeauthoryear{{Croft}}{{Croft}}{2013}]{Croft:2013MNRAS}
{Croft} R. A.~C.,  2013, \mn@doi [Monthly Notices of the Royal Astronomical
  Society] {10.1093/mnras/stt1223}, \href
  {https://ui.adsabs.harvard.edu/abs/2013MNRAS.434.3008C} {434, 3008}

\bibitem[\protect\citeauthoryear{{Crosta}, {Giammaria}, {Lattanzi}  \&
  {Poggio}}{{Crosta} et~al.}{2020}]{Crosta:2020}
{Crosta} M.,  {Giammaria} M.,  {Lattanzi} M.~G.,   {Poggio} E.,  2020, \mn@doi
  [\mnras] {10.1093/mnras/staa1511}, \href
  {https://ui.adsabs.harvard.edu/abs/2020MNRAS.496.2107C} {496, 2107}

\bibitem[\protect\citeauthoryear{Cuesta-Lazaro, Quera-Bofarull, Reischke  \&
  Sch\"afer}{Cuesta-Lazaro et~al.}{2018}]{Cuesta-Lazaro:2018}
Cuesta-Lazaro C.,  Quera-Bofarull A.,  Reischke R.,   Sch\"afer B.~M.,  2018,
  \mn@doi [Mon. Not. Roy. Astron. Soc.] {10.1093/mnras/sty672}, 477, 741

\bibitem[\protect\citeauthoryear{Cusin, Tansella  \& Durrer}{Cusin
  et~al.}{2017}]{cusin_vorticity_2017}
Cusin G.,  Tansella V.,   Durrer R.,  2017, \mn@doi [Physical Review D]
  {10.1103/PhysRevD.95.063527}, 95, 063527

\bibitem[\protect\citeauthoryear{Durrer \& Tansella}{Durrer \&
  Tansella}{2016}]{durrer_vector_2016}
Durrer R.,  Tansella V.,  2016, \mn@doi [Journal of Cosmology and Astroparticle
  Physics] {10.1088/1475-7516/2016/07/037}, 2016, 037

\bibitem[\protect\citeauthoryear{Everitt et~al.,}{Everitt
  et~al.}{2011}]{GravityProbeB:2011}
Everitt C. W.~F.,  et~al., 2011, \mn@doi [Phys. Rev. Lett.]
  {10.1103/PhysRevLett.106.221101}, 106, 221101

\bibitem[\protect\citeauthoryear{Giblin, Mertens  \& Starkman}{Giblin
  et~al.}{2017}]{Giblin:2017juu}
Giblin J.~T.,  Mertens J.~B.,   Starkman G.~D.,  2017, \mn@doi [Class. Quant.
  Grav.] {10.1088/1361-6382/aa8af9}, 34, 214001

\bibitem[\protect\citeauthoryear{Giblin, Mertens, Starkman  \& Tian}{Giblin
  et~al.}{2019}]{Giblin:2018ndw}
Giblin J.~T.,  Mertens J.~B.,  Starkman G.~D.,   Tian C.,  2019, \mn@doi [Phys.
  Rev. D] {10.1103/PhysRevD.99.023527}, 99, 023527

\bibitem[\protect\citeauthoryear{Gressel, Bonvin, Bruni  \& Bacon}{Gressel
  et~al.}{2019}]{Gressel:2019jxw}
Gressel H.~A.,  Bonvin C.,  Bruni M.,   Bacon D.,  2019, \mn@doi [JCAP]
  {10.1088/1475-7516/2019/05/045}, 05, 045

\bibitem[\protect\citeauthoryear{Hahn, Angulo  \& Abel}{Hahn
  et~al.}{2015}]{hahn_properties_2015}
Hahn O.,  Angulo R.~E.,   Abel T.,  2015, \mn@doi [Monthly Notices of the Royal
  Astronomical Society] {10.1093/mnras/stv2179}, 454, 3920

\bibitem[\protect\citeauthoryear{Hand, Feng, Beutler, Li, Modi, Seljak  \&
  Slepian}{Hand et~al.}{2018}]{nbodykit:2018}
Hand N.,  Feng Y.,  Beutler F.,  Li Y.,  Modi C.,  Seljak U.,   Slepian Z.,
  2018, \mn@doi [Astron. J.] {10.3847/1538-3881/aadae0}, 156, 160

\bibitem[\protect\citeauthoryear{He, Li  \& Hawken}{He
  et~al.}{2015}]{He:2015bua}
He J.-h.,  Li B.,   Hawken A.~J.,  2015, \mn@doi [Phys. Rev. D]
  {10.1103/PhysRevD.92.103508}, 92, 103508

\bibitem[\protect\citeauthoryear{Hockney \& Eastwood}{Hockney \&
  Eastwood}{1988}]{CIC-book}
Hockney R.~W.,  Eastwood J.~W.,  Inc., Bristol, PA, USA, 1988, Computer
  Simulation using particles.
Taylor \& Francis

\bibitem[\protect\citeauthoryear{Jelic-Cizmek, Lepori, Adamek  \&
  Durrer}{Jelic-Cizmek et~al.}{2018}]{jelic-cizmek_generation_2018}
Jelic-Cizmek G.,  Lepori F.,  Adamek J.,   Durrer R.,  2018, \mn@doi [Journal
  of Cosmology and Astroparticle Physics] {10.1088/1475-7516/2018/09/006},
  2018, 006

\bibitem[\protect\citeauthoryear{Jolicoeur, Allahyari, Clarkson, Larena, Umeh
  \& Maartens}{Jolicoeur et~al.}{2019}]{jolicoeur_imprints_2019}
Jolicoeur S.,  Allahyari A.,  Clarkson C.,  Larena J.,  Umeh O.,   Maartens R.,
   2019, \mn@doi [Journal of Cosmology and Astroparticle Physics]
  {10.1088/1475-7516/2019/03/004}, 2019, 004

\bibitem[\protect\citeauthoryear{{Lepori}, {Adamek}, {Durrer}, {Clarkson}  \&
  {Coates}}{{Lepori} et~al.}{2020}]{Lepori:2020ifz}
{Lepori} F.,  {Adamek} J.,  {Durrer} R.,  {Clarkson} C.,   {Coates} L.,  2020,
  \mn@doi [\mnras] {10.1093/mnras/staa2024}, \href
  {https://ui.adsabs.harvard.edu/abs/2020MNRAS.497.2078L} {497, 2078}

\bibitem[\protect\citeauthoryear{Lewis, Challinor  \& Lasenby}{Lewis
  et~al.}{2000}]{CAMB}
Lewis A.,  Challinor A.,   Lasenby A.,  2000, \mn@doi [Astrophys. J.]
  {10.1086/309179}, 538, 473

\bibitem[\protect\citeauthoryear{Li, Zhao, Teyssier  \& Koyama}{Li
  et~al.}{2012}]{Li2012:ECOSMOG}
Li B.,  Zhao G.-B.,  Teyssier R.,   Koyama K.,  2012, \mn@doi [JCAP]
  {10.1088/1475-7516/2012/01/051}, 1201, 051

\bibitem[\protect\citeauthoryear{{Li}, {Zhao}  \& {Koyama}}{{Li}
  et~al.}{2013}]{Li2013:ECOSMOG}
{Li} B.,  {Zhao} G.-B.,   {Koyama} K.,  2013, \mn@doi [JCAP]
  {10.1088/1475-7516/2013/05/023}, \href
  {https://ui.adsabs.harvard.edu/abs/2013JCAP...05..023L} {2013, 023}

\bibitem[\protect\citeauthoryear{Linder}{Linder}{2005}]{Linder-growth}
Linder E.~V.,  2005, \mn@doi [Phys. Rev. D] {10.1103/PhysRevD.72.043529}, 72,
  043529

\bibitem[\protect\citeauthoryear{Lu, Ananda, Clarkson  \& Maartens}{Lu
  et~al.}{2009}]{Lu:2008ju}
Lu T. H.-C.,  Ananda K.,  Clarkson C.,   Maartens R.,  2009, \mn@doi [JCAP]
  {10.1088/1475-7516/2009/02/023}, 0902, 023

\bibitem[\protect\citeauthoryear{Macpherson, Lasky  \& Price}{Macpherson
  et~al.}{2017}]{Macpherson:2016ict}
Macpherson H.~J.,  Lasky P.~D.,   Price D.~J.,  2017, \mn@doi [Phys. Rev.]
  {10.1103/PhysRevD.95.064028}, D95, 064028

\bibitem[\protect\citeauthoryear{Matarrese, Mollerach  \& Bruni}{Matarrese
  et~al.}{1998a}]{matarrese_relativistic_1998}
Matarrese S.,  Mollerach S.,   Bruni M.,  1998a, \mn@doi [Physical Review D]
  {10.1103/PhysRevD.58.043504}, 58, 043504

\bibitem[\protect\citeauthoryear{Matarrese, Mollerach  \& Bruni}{Matarrese
  et~al.}{1998b}]{Matarrese:1997ay}
Matarrese S.,  Mollerach S.,   Bruni M.,  1998b, \mn@doi [Phys. Rev.]
  {10.1103/PhysRevD.58.043504}, D58, 043504

\bibitem[\protect\citeauthoryear{McDonald}{McDonald}{2009}]{McDonald:2009}
McDonald P.,  2009, \mn@doi [Journal of Cosmology and Astroparticle Physics]
  {10.1088/1475-7516/2009/11/026}, 2009, 026

\bibitem[\protect\citeauthoryear{Mertens, Giblin  \& Starkman}{Mertens
  et~al.}{2016}]{Mertens:2015ttp}
Mertens J.~B.,  Giblin J.~T.,   Starkman G.~D.,  2016, \mn@doi [Phys. Rev.]
  {10.1103/PhysRevD.93.124059}, D93, 124059

\bibitem[\protect\citeauthoryear{Milillo, Bertacca, Bruni  \& Maselli}{Milillo
  et~al.}{2015}]{Milillo:2015cva}
Milillo I.,  Bertacca D.,  Bruni M.,   Maselli A.,  2015, \mn@doi [Phys. Rev.
  D] {10.1103/PhysRevD.92.023519}, 92, 023519

\bibitem[\protect\citeauthoryear{Navarro, Frenk  \& White}{Navarro
  et~al.}{1996}]{Navarro:1995iw}
Navarro J.~F.,  Frenk C.~S.,   White S.~D.,  1996, \mn@doi [Astrophys. J.]
  {10.1086/177173}, 462, 563

\bibitem[\protect\citeauthoryear{Pueblas \& Scoccimarro}{Pueblas \&
  Scoccimarro}{2009}]{pueblas_generation_2009}
Pueblas S.,  Scoccimarro R.,  2009, \mn@doi [Physical Review D]
  {10.1103/PhysRevD.80.043504}, 80, 043504

\bibitem[\protect\citeauthoryear{Reverdy}{Reverdy}{2014}]{reverdy-thesis}
Reverdy V.,  2014, PhD Thesis.
Laboratoire Univers et T\'heories

\bibitem[\protect\citeauthoryear{Saga, Yamauchi  \& Ichiki}{Saga
  et~al.}{2015}]{Saga:2015apa}
Saga S.,  Yamauchi D.,   Ichiki K.,  2015, \mn@doi [Phys. Rev. D]
  {10.1103/PhysRevD.92.063533}, 92, 063533

\bibitem[\protect\citeauthoryear{{Sereno}}{{Sereno}}{2002}]{Sereno:2002}
{Sereno} M.,  2002, \mn@doi [Physics Letters A]
  {10.1016/S0375-9601(02)01361-0}, \href
  {https://ui.adsabs.harvard.edu/abs/2002PhLA..305....7S} {305, 7}

\bibitem[\protect\citeauthoryear{{Sereno}}{{Sereno}}{2003}]{Sereno:2003}
{Sereno} M.,  2003, \mn@doi [\prd] {10.1103/PhysRevD.67.064007}, \href
  {https://ui.adsabs.harvard.edu/abs/2003PhRvD..67f4007S} {67, 064007}

\bibitem[\protect\citeauthoryear{{Sereno}}{{Sereno}}{2007}]{Sereno:2007}
{Sereno} M.,  2007, \mn@doi [\mnras] {10.1111/j.1365-2966.2007.12126.x}, \href
  {https://ui.adsabs.harvard.edu/abs/2007MNRAS.380.1023S} {380, 1023}

\bibitem[\protect\citeauthoryear{Shibata}{Shibata}{1999}]{Shibata:1999va}
Shibata M.,  1999, \mn@doi [Prog. Theor. Phys.] {10.1143/PTP.101.251}, 101, 251

\bibitem[\protect\citeauthoryear{Shibata \& Sasaki}{Shibata \&
  Sasaki}{1999}]{Shibata:1999zs}
Shibata M.,  Sasaki M.,  1999, \mn@doi [Phys. Rev.]
  {10.1103/PhysRevD.60.084002}, D60, 084002

\bibitem[\protect\citeauthoryear{Smarr \& York}{Smarr \&
  York}{1978a}]{Smarr1978:MDC}
Smarr L.,  York J.~W.,  1978a, \mn@doi [Phys. Rev. D]
  {10.1103/PhysRevD.17.1945}, 17, 1945

\bibitem[\protect\citeauthoryear{Smarr \& York}{Smarr \&
  York}{1978b}]{Smarr-York:MEC-1978}
Smarr L.,  York J.~W.,  1978b, \mn@doi [Phys. Rev. D]
  {10.1103/PhysRevD.17.2529}, 17, 2529

\bibitem[\protect\citeauthoryear{{Tang}, {Zhang}, {Luo}, {Li}, {Cai}  \&
  {Pi}}{{Tang} et~al.}{2020}]{Tang_gravitomagnetic:2020}
{Tang} C.,  {Zhang} P.,  {Luo} W.,  {Li} N.,  {Cai} Y.-F.,   {Pi} S.,  2020,
  arXiv e-prints, \href {https://ui.adsabs.harvard.edu/abs/2020arXiv200912011T}
  {p. arXiv:2009.12011}

\bibitem[\protect\citeauthoryear{Tansella, Bonvin, Cusin, Durrer, Kunz  \&
  Sawicki}{Tansella et~al.}{2018}]{tansella_redshift-space_2018}
Tansella V.,  Bonvin C.,  Cusin G.,  Durrer R.,  Kunz M.,   Sawicki I.,  2018,
  \mn@doi [Physical Review D] {10.1103/PhysRevD.98.103515}, 98, 103515

\bibitem[\protect\citeauthoryear{Taylor \& Jagannathan}{Taylor \&
  Jagannathan}{2016}]{Taylor_2016}
Taylor A.~R.,  Jagannathan P.,  2016, \mn@doi [Monthly Notices of the Royal
  Astronomical Society: Letters] {10.1093/mnrasl/slw038}, 459, L36–L40

\bibitem[\protect\citeauthoryear{Teyssier}{Teyssier}{2002}]{Teyssier:ramses}
Teyssier R.,  2002, \mn@doi [Astron. Astrophys.] {10.1051/0004-6361:20011817},
  385, 337

\bibitem[\protect\citeauthoryear{Thomas, Bruni  \& Wands}{Thomas
  et~al.}{2015a}]{Thomas:2014aga}
Thomas D.~B.,  Bruni M.,   Wands D.,  2015a, \mn@doi [JCAP]
  {10.1088/1475-7516/2015/9/021}, 09, 021

\bibitem[\protect\citeauthoryear{Thomas, Bruni  \& Wands}{Thomas
  et~al.}{2015b}]{thomas_fully_2015}
Thomas D.~B.,  Bruni M.,   Wands D.,  2015b, \mn@doi [Monthly Notices of the
  Royal Astronomical Society] {10.1093/mnras/stv1390}, 452, 1727

\bibitem[\protect\citeauthoryear{Thomas, Bruni, Koyama, Li  \& Zhao}{Thomas
  et~al.}{2015c}]{thomas_fr_2015}
Thomas D.~B.,  Bruni M.,  Koyama K.,  Li B.,   Zhao G.-B.,  2015c, \mn@doi
  [Journal of Cosmology and Astroparticle Physics]
  {10.1088/1475-7516/2015/07/051}, 2015, 051

\makeatother
\end{thebibliography}



\appendix

\section{Comparison of power spectrum calculation methods}
\label{appendix}

In Section~\ref{sec:spectra}, the power spectrum of density, velocity and vorticity has been measured from particle-type data using \dtfe~and {\sc nbodykit}, while the spectrum of the scalar and vector potentials has been measured using a different code that is able to read their values calculated and stored by \gramses\ in cells of hierarchical AMR meshes and interpolate them to a regular grid for the power spectrum measurement. We call this method the `AMR-FFT' method, which was introduced in \citet{He:2015bua}, where more details can be found. An alternative to using this AMR-FFT method to calculate the power spectrum of the potentials is by writing their values with \gramses\ at the particles' positions rather than in AMR cells, so that \dtfe{} can be used to read such `particle-type' data and interpolate this to a regular grid, where {\sc nbodykit} can be applied to measure the spectrum. We call this method `{\sc dtfe}+{\sc nbodykit}'.

\begin{figure*}
    \centering
    \includegraphics[width=\linewidth]{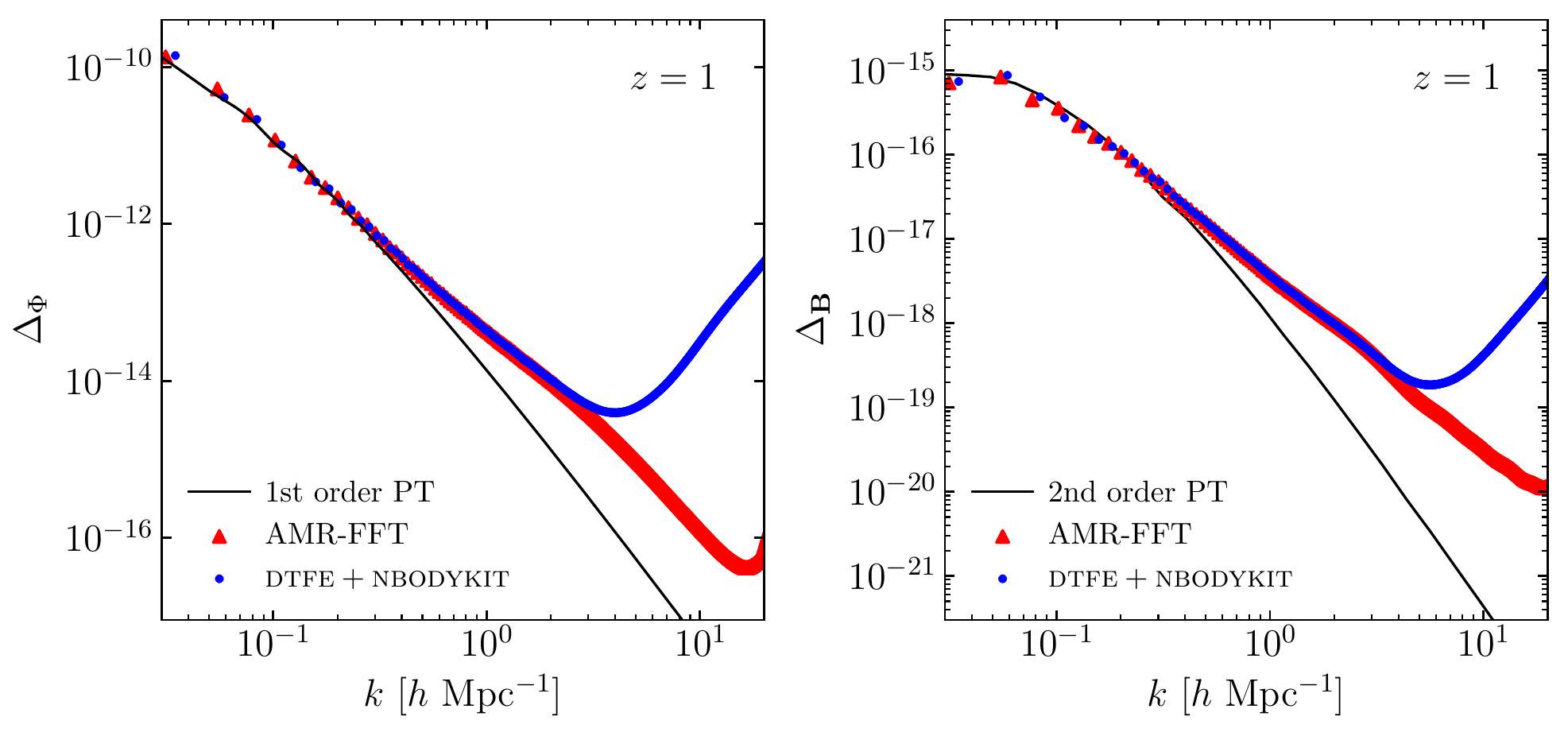}
    \caption{(Colour Online) Comparison of the power spectra of the scalar and vector potentials measured with the AMR-FFT method, and {\sc nbodykit} combined with \dtfe. In both methods the grid size used for the FFT is $2048^3$, and is equal to the tessellation grid size used in \dtfe. Both panels show the dimensionless power spectrum $\Delta_{\bf}(k)=k^3P_{\bf}(k)/(2\pi^2)$ of the respective field. {\it Left}: The dimensionless power spectrum of the of the scalar gravitational potential $\Phi$ defined as the fully nonlinear perturbation to the lapse function, i.e., $\Phi\equiv\alpha-1$. The solid line represents the first-order perturbation theory prediction of the Bardeen potential from \camb. 
    {\it Right}: The dimensionless power spectrum of the vector potential ${\bf B}$. The solid line corresponds to the second-order perturbation theory result from Eq.~\eqref{eq:delta-beta-v-convolution}. All results are at $z=1$. 
    } 
    \label{fig:Pk-methods-comp}
\end{figure*}

Figure~\ref{fig:Pk-methods-comp} shows the dimensionless power spectra at $z=1$ of the scalar potential $\Phi$ (left panel) and the vector potential spectrum (right panel), measured by these two methods, where solid lines represent the perturbation-theory predictions. In both methods the FFT grid size is $2048^3$, as is the tessellation grid size used for \dtfe. We find that both methods have good agreement on large scales, specially at $k\gtrsim0.1\invMpch$, where the effect of cosmic variance is not present. However, in the region $k\gtrsim3\invMpch$ the AMR-FFT method has better performance than \dtfe+{\sc nbodykit} which blows up. This is because the AMR-FFT method can reach higher resolution by using the potential information in the AMR cells, and because \dtfe{} does a volume weighted average of the field which smears out small-scale features. Therefore, the spectrum of the scalar and vector potentials from the simulation shown in Fig.~\ref{fig:shift-vector-power} are measured by the AMR-FFT method, which yields robust results up to $k\sim15\invMpch$.

\bsp	
\label{lastpage}
\end{document}